\documentclass[showpacs,amssymb,floatfix,prb,notitlepage]{revtex4-1}

\usepackage{amssymb,amsfonts,bm}
\usepackage{graphics,graphicx}
\usepackage{sidecap} 

\begin{document}

\title{Critical phenomena and phase sequence in classical bilayer Wigner crystal at 
zero temperature}

\author{Ladislav \v{S}amaj}
\altaffiliation[On leave from ]
{Institute of Physics, Slovak Academy of Sciences, Bratislava, Slovakia}
\author{Emmanuel Trizac}
\affiliation{Laboratoire de Physique Th\'eorique et Mod\`eles Statistiques, 
UMR CNRS 8626, Universit\'e Paris-Sud, 91405 Orsay, France}

\begin{abstract}
We study the ground-state properties of a system of identical classical 
Coulombic point particles, evenly distributed between two equivalently charged 
parallel plates at distance $d$; the system as a whole is electroneutral.
It was previously shown that upon increasing $d$ from $0$ to $\infty$, five different 
structures of the bilayer Wigner crystal become energetically favored,
starting from a hexagonal lattice (phase I, $d=0$) and 
ending at a staggered hexagonal lattice (phase V, $d\to\infty$).
In this paper, we derive new series representations of the ground-state
energy for all five bilayer structures.
The derivation is based on a sequence of transformations for lattice
sums of Coulomb two-particle potentials plus the neutralizing background, 
having their origin in the general theory of Jacobi theta functions.
The new series provide convenient starting points for both analytical and
numerical progress.
Its convergence properties are indeed excellent: Truncation 
at the fourth term determines in general the energy correctly 
up to 17 decimal digits. 
The accurate series representations are used to improve the specification
of transition points between the phases and to solve a controversy in
previous studies.
In particular, it is shown both analytically and numerically that 
the hexagonal phase I is stable only at $d=0$, and not in a finite interval
of small distances between the plates as was anticipated before.
The expansions of the structure energies around second-order transition 
points can be done analytically, which enables us to show that
the critical behavior is of the Ginzburg-Landau type,
with a mean-field critical index $\beta=1/2$ for the growth of
the order parameters.   
\end{abstract}


\date{\today} 

\maketitle

\section{Introduction} \label{sec:intro}
Classical charged particles, confined in a 
two-dimensional (2D) layer and interacting via the usual three-dimensional 
Coulomb potential, exhibit a crystallization into a Wigner hexagonal structure, 
when kinetic energy is small compared to potential energy.\cite{Wi34,GrAd79}
We shall be interested here in bilayer systems, that 
describe several properties of real 
physical systems in condensed and soft matter, such as 
semiconductors,\cite{junctions} quantum dots,\cite{ImMA96} 
boron nitride,\cite{boron} laser-cooled trapped ion plasmas,\cite{Mitc98}
dusty plasmas\cite{dusty} and colloids.\cite{colloids}
For a recent review of numerical methods for quasi-2D systems with
long-range interactions, see Ref.\cite{Mazars11}
In addition, the creation of
a bilayer Wigner crystal on two charged plates at some distance is 
of primary importance in the study of ``anomalous'' strong-coupling 
effects such as like-charge attraction or 
overcharging.\cite{LaLP00,Grosberg02,Levin02,Naji05,Samaj11,Samaj12}

\begin{SCfigure}[1.5][t]
\includegraphics[height=10cm,clip]{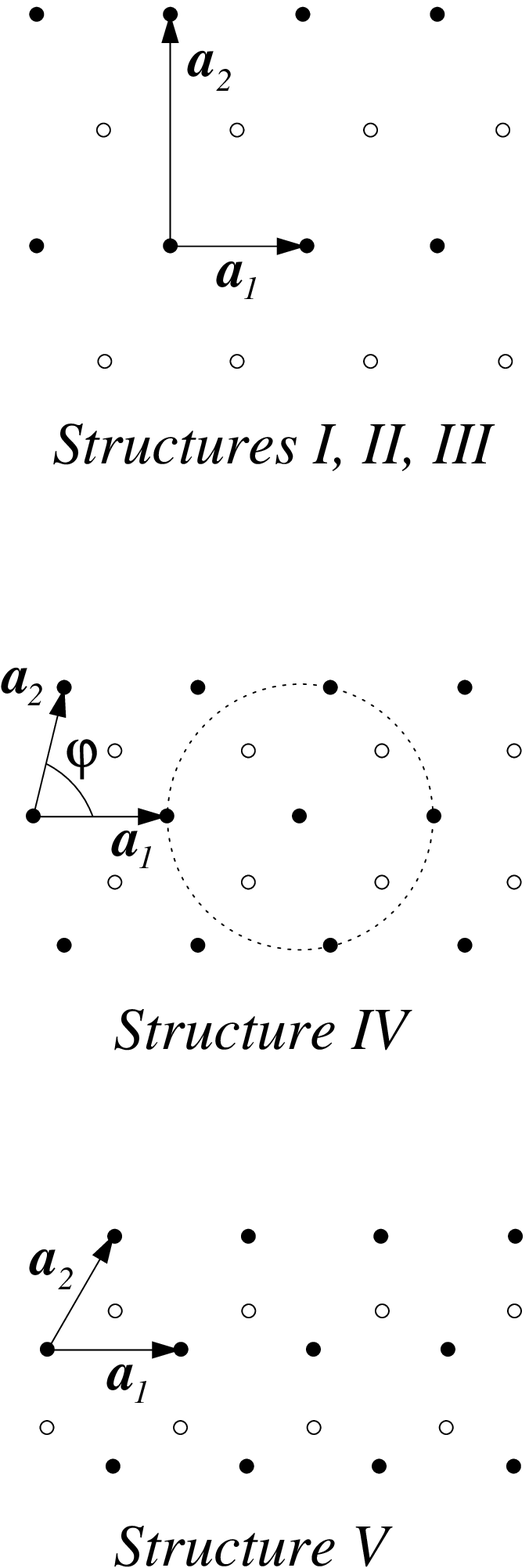}
\caption{Ground-state structures I--V of counter-ions on two parallel 
equivalently and homogeneously charged plates. 
Open and filled symbols correspond to particle positions on 
the opposite surfaces.
The primitive translation vectors of the Bravais lattice on one of
the surfaces are denoted by $\bm{a}_1$ and $\bm{a}_2$.
For structures I, II and III, we define the aspect ratio as
$\Delta=|\bm{a}_1|/|\bm{a}_2|$, so that $\Delta=\sqrt{3}$ 
with structure I, $1 < \Delta < \sqrt{3}$ for structure II
and $\Delta=1$ for structure III. The dashed circle for structure
IV is a guide to the eye, for identifying those points that are
equidistant to the ion in the circle center.
For a more detailed description of the structures, see the text.}
\label{fig:Structures}
\end{SCfigure}

In this paper, we study the ground-state properties of a classical 
one-component plasma of identical Coulombic particles of the charge $-e$, 
evenly distributed between two plates of the same homogeneous
fixed charge density 
$\sigma e$ which are at distance $d$. 
The total surface density of the particles is $n$, the particle density
in each layer is $n_l=n/2$.
The overall electroneutrality of the system is ensured by the
condition $n_l=\sigma$.
The phase diagram of the system at temperature $T=0$ is determined
by a single dimensionless parameter $\eta=d\sqrt{n/2}=d\sqrt{\sigma}$.
By comparing the static energy of various lattices, five distinct
phases were detected to be stable (providing global minimum of the energy) 
in different ranges of $\eta$.\cite{Falk94,EsKa95,Goldoni96,ScSP99,
Weis01,Messina03,Lobaskin07}
In order of increasing $\eta$, these phases are:
a hexagonal lattice (I) for $\eta\in[0,\eta_1^c]$, 
a staggered rectangular lattice (II) for $\eta\in[\eta_1^c,\eta_2^c]$, 
a staggered square lattice (III) for $\eta\in[\eta_2^c,\eta_3^c]$, 
a staggered rhombic lattice (IV) for $\eta\in[\eta_3^c,\eta_4^c]$ and 
a staggered hexagonal lattice (V) for $\eta\in[\eta_4^c,\infty]$;
although we use an index $c$ in $\eta^c$, the transition point $\eta^c$ 
from one structure to the other is not necessarily a critical point.
The structures are pictured in Fig. \ref{fig:Structures}. 
The different symbols correspond to particle positions on the opposite 
surfaces.
The primitive translation vectors of the Bravais lattice on one of
the surfaces are denoted by $\bm{a}_1$ and $\bm{a}_2$.

The ground-state structures I, III and V are ``rigid'', i.e. they have 
fixed ($\eta$-independent) primary cells within their region 
of stability. 
The structures II and IV are ``soft'', i.e. the  shape of their 
primary cells is varying with increasing $\eta$, within their region
of stability.
We now outline the basic characteristics of the structures.\cite{Falk94,EsKa95,Goldoni96,ScSP99,
Weis01,Messina03,Lobaskin07}
\begin{itemize}
\item
{\bf Structures I, II and III:}
Within one single layer, the structure corresponds to a rectangular lattice
with the aspect ratio $\Delta=\vert \bm{a}_2\vert/\vert \bm{a}_1\vert$.
The equivalent structures on the two layers are shifted with respect 
to one another by a half period, i.e. by $(\bm{a}_1+\bm{a}_2)/2$. 

Structure I with $\Delta=\sqrt{3}$ arises naturally in the simple limit 
$\eta\to 0$, where the bilayer structure reduces to a single layer,
which is known to crystallize in a hexagonal (equilateral triangular) 
lattice.\cite{Bonsall,Jagla}
An open question is whether phase I (with the fixed aspect ratio
$\Delta=\sqrt{3}$) exists only at $\eta=0$ or is stable also in 
a finite interval $[0,\eta_1^c]$ with some $\eta_1^c>0$.
Some numerical calculations indicate very small, but nonzero values of
$\eta_1^c=0.006$ (Ewald technique\cite{Goldoni96}) and 
$0.028$ (Monte Carlo simulations\cite{Weis01}). On the other hand,
another study for Yukawa bilayers in the limit of infinite screening
length indicates that $\eta_1^c=0$, so that a buckled phase of type II preempts
structure I when $\eta$ is small but non vanishing.\cite{Messina03}

Structure II continuously interpolates between
the rigid structures I and III.
The value of the aspect ratio $\Delta$ then changes smoothly
from $\sqrt{3}$ at $\eta_1^c$ (phase I) to $1$ at the transition point 
$\eta_2^c$ to phase III.
It is not clear whether or not $\eta_1^c$, zero or nonzero, is a standard
transition point between phases I and II.
The transition between phases II and III is continuous (of second order).
\item
The structure IV is characterized by an angle $\theta$ between primitive 
cell vectors $\bm{a}_1$ and $\bm{a}_2$ of the same length $a$.
Increasing $\eta$, the angle $\varphi$ changes continuously from 
$\pi/2$ at $\eta_3^c$ (continuous transition between phases III and IV) 
up to $\eta_4^c$, where it drops to $\pi/3$.
Together with an additional shift between the sub-lattices on the two layers, 
this corresponds to a discontinuous (first order) transition to phase V. 
\item
The presence of  structure V is expected for large enough
$\eta\ge \eta_4^c$: 
At large separation between the layers, the two particle sub-lattices 
are only weakly coupled and so two staggered hexagonal lattices form 
the stable structure.
\end{itemize}

In this paper, we derive new series representations of the energy for
all five bilayer structures.
The derivation is based on a sequence of transformations for lattice
sums of Coulomb potentials plus the neutralizing background, having
their origin in the general theory of Jacobi theta functions.\cite{Abra}
The series has excellent convergence properties, which is 
convenient for numerical investigations, but is also 
conducive to analytical progress.
It will be used to improve the specification
of transition points between the phases and to solve the aforementioned controversy 
concerning the stability of phase I.
In particular, it is shown both analytically and numerically that 
phase I is realized only at $\eta=0$, i.e. $\eta_1^c=0$.
The expansions of the structure energies around second-order transition 
points can be done analytically which enables us to derive the critical 
exponents at phase transition points. 
The critical behavior is of the Ginzburg-Landau type,\cite{GL} with the mean-field 
critical index $\beta=1/2$ for the growth of the order parameters.   
A preliminary account of this work has appeared in Ref.\cite{EPL}

The paper is organized as follows.
Sec. \ref{sec:123} is devoted to a detailed derivation of the
series representation of the energy for structures I-III.
The existence of phase I at $\eta=0$ only is established 
analytically and illustrated numerically.
The second-order phase transition between phases II and III is then
described.
The energy of structure IV and the second-order phase transition
between phases III and IV are treated in Sec. \ref{sec:4}.
Sec. \ref{sec:5} deals with structure V and the first-order
phase transition between phases IV and V, while our conclusions are finally
presented in Sec.
\ref{sec:conclusion}.

\section{Phases I-III} \label{sec:123}
Structures I, II and III are treated on equal footings by considering 
the general case of structure II, see Fig. \ref{fig:Structures}.
For one single layer, the 2D lattice points are indexed by 
$j \bm{a}_1 + k \bm{a}_2$, where $j,k$ are any two integers 
(positive, negative or zero) and
\begin{equation}
\bm{a}_1 = a (1,0) , \qquad \bm{a}_2 = a (0,\Delta) \qquad
{\rm with}\ a=\frac{1}{\sqrt{\sigma\Delta}}
\end{equation}
are the primitive translation vectors of the Bravais lattice.
The lattice spacing $a$ is determined by the electroneutrality condition 
$n_l=\sigma$ with the one-layer particle density $n_l=1/(\Delta a^2)$. 
The aspect ratio $\Delta$ is a continuous parameter in the interval 
$[1,\sqrt{3}]$; as was already mentioned, the limiting cases $\sqrt{3}$ 
and $1$ correspond to the phases I and III, respectively.

\subsection{Energy of phases I-III} 
The dielectric constant of the medium is set to unity for simplicity,
and we start by a preliminary remark, valid for all phases.
Our goal is to compute the total electrostatic energy, 
including particle-particle, particle-plate,
and plate-plate interactions.  
The latter two contributions per unit surface
can be derived straightforwardly, and respectively read $4 \pi \sigma^2 e^2 d$
and $-2 \pi \sigma^2 e^2 d$. The sum of both, $2 \pi \sigma^2 e^2 d$,
thus gives $1/2$ of the particle-plate energy, and this is why in the
subsequent analysis, we shall add to the non trivial particle-particle
energy one half of the particle-plate energy (also referred
to as the particle-background term).
The resulting sum provides the full energy of the system.

The energy per particle $E$ of the bilayer system consists of the
intralayer and interlayer contributions,
\begin{equation}
E = E_{\rm intra} + E_{\rm inter} .
\end{equation}
We first consider the intralayer contribution.
It is well known that lattice sums involving the pair Coulomb interactions
exhibit infinities which are canceled exactly by the neutralizing
background term.\cite{Mazars11,dLPS80}
To maintain mathematical rigor, we first restrict ourselves to 
a disk of finite radius $R$ around a reference particle localized
at the origin $(0,0)$.
The interaction energy due to the discrete Wigner crystal is given by
\begin{equation} \label{contr1}
\frac{e^2}{2a} \sum_{j,k\atop (j,k)\ne (0,0)}
\frac{1}{\sqrt{j^2 + k^2 \Delta^2}} , \qquad
j^2 +k^2 \Delta^2 \le \left( \frac{R}{a} \right)^2 . 
\end{equation}
Hereinafter, the omission of the lower and upper values for integer
indices $j,k$ automatically means a summation from $-\infty$ to $\infty$.
The interaction of the reference particle with the 2D charge background 
in the disk is expressed as
\begin{equation} \label{contr2}
- \frac{\sigma e^2}{2} \int_0^R d^2{\bf r} \frac{1}{\vert{\bf r}\vert}
=  - \frac{e^2}{2a\Delta} \int_0^{R/a} dr 2\pi r \frac{1}{r} .
\end{equation}
$E_{\rm intra}$ is the sum of (\ref{contr1}) plus (\ref{contr2}).
We intend to rewrite $E_{\rm intra}$ by using the gamma identity
\begin{equation} \label{gamma}
\frac{1}{\sqrt{z}} = \frac{1}{\sqrt{\pi}} \int_0^{\infty} 
\frac{dt}{\sqrt{t}} e^{-zt} , \qquad z>0 ,
\end{equation}
a common procedure in the field.\cite{Mazars11,dLPS80}
Each term $1/(j^2 + k^2 \Delta^2)^{1/2}$ in
(\ref{contr1}) can consequently be written as
\begin{equation}
\frac{1}{\sqrt{j^2 + k^2 \Delta^2}} = \frac{1}{\sqrt{\pi}} \int_0^{\infty} 
\frac{dt}{\sqrt{t}} e^{-t j^2} e^{-t\Delta^2 k^2} .
\end{equation}
As concerns the background contribution (\ref{contr2}), the application of 
the identity (\ref{gamma}) to the term $1/r = 1/\sqrt{r^2}$ under integration
leads to
\begin{equation}
\int_0^{R/a} dr 2\pi r \frac{1}{r} = \int_0^{R/a} dr 2\pi r
\frac{1}{\sqrt{\pi}} \int_0^{\infty} \frac{dt}{\sqrt{t}} e^{-t r^2} = 
\frac{1}{\sqrt{\pi}} \int_0^{\infty} 
\frac{dt}{\sqrt{t}} \frac{\pi}{t}\left[ 1 - e^{-t(R/a)^2} \right] . 
\end{equation}
Altogether, we get
\begin{equation}
E_{\rm intra} = \frac{e^2}{2a\sqrt{\pi}} \int_0^{\infty} \frac{dt}{\sqrt{t}} 
\Bigg\{ \sum_{j,k} e^{-t j^2} e^{-t\Delta^2 k^2} 
- 1 - \frac{\pi}{t\Delta} \left[ 1 - e^{-t(R/a)^2} \right] \Bigg\} ,
\qquad j^2 +k^2 \Delta^2 \le \left( \frac{R}{a} \right)^2 . 
\end{equation}
Here, the subtraction of unity is due to the absence of the term
$(j,k)=(0,0)$ in the sum (\ref{contr1}).
Having all contributions under the same integration, we are allowed to take 
the limit $R/a\to \infty$, which removes the exponentially small term
$\exp[-t(R/a)^2]$ and the disk constraint for lattice indices.
Using the definition of the Jacobi theta function with zero 
argument\cite{Gradshteyn} $\theta_3(q)=\sum_j q^{j^2}$ 
and making the substitution $t \Delta\to t$, we end up with the result
\begin{equation} \label{theta3}
\frac{E_{\rm intra}}{e^2\sqrt{n}} = \frac{1}{2^{3/2}\sqrt{\pi}}
\int_0^{\infty} \frac{dt}{\sqrt{t}} 
\left[ \theta_3(e^{-t\Delta}) \theta_3(e^{-t/\Delta}) - 1 - \frac{\pi}{t} \right] .
\end{equation}

We shall repeatedly use the Poisson summation formula
\begin{equation} \label{Poisson}
\sum_{j=-\infty}^{\infty} e^{-(j+\phi)^2 t} = \sqrt{\frac{\pi}{t}}
\sum_{j=-\infty}^{\infty} e^{2\pi ij\phi}e^{-(\pi j)^2/t} .
\end{equation} 
The asymptotic behaviors
\begin{equation}
\theta_3(e^{-t}) \mathop{\sim}_{t\to 0} \sqrt{\frac{\pi}{t}} \left( 1 +
2 e^{-\pi^2/t} + \cdots \right) , \qquad 
\theta_3(e^{-t}) \mathop{\sim}_{t\to\infty} 1 + 2 e^{-t} + \cdots
\end{equation}
follow immediately.
We see that the background charge contribution $-\pi/t$ correctly cancels 
the $t\to 0$ singularity of the product of two $\theta_3$ functions inside 
the square bracket in (\ref{theta3}) and the integral converges. 

The Wigner lattices on the opposite layers are shifted with respect to
one another by the vector $(\bm{a}_1+\bm{a}_2)/2$, see Fig. \ref{fig:Structures}.
To obtain the interlayer contribution to the energy, we first consider
the disk of radius $R$ around the (perpendicular) image of the reference 
particle on the opposite layer.
The interaction energy of the Wigner crystal is given by
\begin{equation}
\frac{e^2}{2a} \sum_{j,k} \frac{1}{\sqrt{\left( j-\frac{1}{2}\right)^2 + 
\left( k-\frac{1}{2}\right)^2 \Delta^2 +(d/a)^2}} , \qquad
\left( j-\frac{1}{2}\right)^2 +\left( k-\frac{1}{2}\right)^2 \Delta^2 
\le \left( \frac{R}{a} \right)^2.
\end{equation}
The interaction with the background charge is described by
\begin{equation}
- \frac{\sigma e^2}{2} \int_0^R d^2{\bf r} \frac{1}{\vert{\bf r}+{\bf d}\vert}
=  - \frac{e^2}{2a\Delta} \int_0^{R/a} dr 2\pi r \frac{1}{\sqrt{r^2+(d/a)^2}} .
\end{equation}
Proceeding as in the previous case and taking into account that
$d/a = \eta \sqrt{\Delta}$, we find
\begin{equation}
\frac{E_{\rm inter}}{e^2\sqrt{n}} = \frac{1}{2^{3/2}\sqrt{\pi}}
\int_0^{\infty} \frac{dt}{\sqrt{t}} e^{-\eta^2 t}
\left[ \theta_2(e^{-t\Delta}) \theta_2(e^{-t/\Delta}) - \frac{\pi}{t} \right]
\label{eq:14}
\end{equation}
with the Jacobi theta function 
$\theta_2(q)=\sum_j q^{\left(j-\frac{1}{2}\right)^2}$. 
It follows from Eq. (\ref{Poisson}) that 
\begin{equation}
\theta_2(e^{-t}) \mathop{\sim}_{t\to 0} \sqrt{\frac{\pi}{t}} 
\left( 1 - 2 e^{-\pi^2/t} + \cdots \right) , \qquad
\theta_2(e^{-t}) \mathop{\sim}_{t\to\infty} 2 \, e^{-t/4} + \cdots ,
\end{equation}
so that the integral in Eq. (\ref{eq:14}) converges, as it should.

The total energy per particle $E$ reads
\begin{eqnarray} \label{E}
\frac{E(\Delta,\eta)}{e^2\sqrt{n}} = \frac{1}{2^{3/2}\sqrt{\pi}}
\int_0^{\infty} \frac{dt}{\sqrt{t}} \Bigg\{ 
\left[ \theta_3(e^{-t\Delta}) \theta_3(e^{-t/\Delta}) - 1 - \frac{\pi}{t} \right]
+ e^{-\eta^2 t} \left[ \theta_2(e^{-t\Delta}) \theta_2(e^{-t/\Delta}) 
- \frac{\pi}{t} \right] \Bigg\} .
\end{eqnarray}
Note the invariance of $E$ with respect to the transformation
$\Delta\to 1/\Delta$, which is physically clear from the configuration
sketched in Fig \ref{fig:Structures} (label exchange of the two Bravais
vectors $\bm{a}_1$ and $\bm{a}_2$).
From a numerical point of view, there are two ``dangerous'' limits:
$t\to 0$ and $t\to\infty$, that jeopardize accuracy.
To simplify the integral representation of $E$, we split the range of
integration into two parts, $[0,\pi]$ and $[\pi,\infty]$, and transform
the integral over $[\pi,\infty]$ to the one over $[0,\pi]$ by using
the Poisson formula (\ref{Poisson}).
For the term containing the product of two $\theta_3$ functions, 
the procedure reads
\begin{eqnarray}
\int_{\pi}^{\infty} \frac{dt}{\sqrt{t}} 
\left[ \theta_3(e^{-t\Delta}) \theta_3(e^{-t/\Delta}) - 1 - \frac{\pi}{t} \right]
& \equiv & \int_{\pi}^{\infty} \frac{dt}{\sqrt{t}} 
\left[ \sum_j e^{-t j^2\Delta} \sum_k e^{-t k^2/\Delta} - 1 - \frac{\pi}{t} \right]
\nonumber \\
& = & \int_{\pi}^{\infty} \frac{dt}{\sqrt{t}} 
\left[ \frac{\pi}{t} \sum_j e^{-(\pi j)^2/(\Delta t)} 
\sum_k e^{-(\pi k)^2\Delta/t} - 1 - \frac{\pi}{t} \right] \nonumber \\
& = & \int_0^{\pi} \frac{\pi dt'}{(t')^{3/2}} 
\left[ \frac{t'}{\pi} \sum_j e^{-t' j^2/\Delta} 
\sum_k e^{-t' k^2\Delta} - 1 - \frac{t'}{\pi} \right] \nonumber \\
& \equiv & \int_0^{\pi} \frac{dt}{\sqrt{t}} 
\left[ \theta_3(e^{-t\Delta}) \theta_3(e^{-t/\Delta}) - 1 - \frac{\pi}{t} \right] .
\end{eqnarray}
Here, going from the third integral to the fourth one, we applied the
substitution $t'=\pi^2/t$.
Similarly, we have
\begin{eqnarray}
\int_{\pi}^{\infty} \frac{dt}{\sqrt{t}} e^{-\eta^2 t}
\left[ \theta_2(e^{-t\Delta}) \theta_2(e^{-t/\Delta}) - \frac{\pi}{t} \right]
& \equiv & \int_{\pi}^{\infty} \frac{dt}{\sqrt{t}} e^{-\eta^2 t} 
\left[ \sum_j e^{-t \left( j-\frac{1}{2}\right)^2\Delta} 
\sum_k e^{-t \left( k-\frac{1}{2}\right)^2/\Delta} - \frac{\pi}{t} \right]
\nonumber \\
& = & \int_{\pi}^{\infty} \frac{dt}{\sqrt{t}} e^{-\eta^2 t}
\left[ \frac{\pi}{t} \sum_j (-1)^j e^{-(\pi j)^2/(\Delta t)} 
\sum_k (-1)^k e^{-(\pi k)^2\Delta/t} - \frac{\pi}{t} \right] \nonumber \\
& = & \int_0^{\pi} \frac{\pi dt'}{(t')^{3/2}} e^{-(\pi\eta)^2/t'}
\left[ \frac{t'}{\pi} \sum_j (-1)^j e^{-t' j^2/\Delta} 
\sum_k (-1)^k e^{-t' k^2\Delta} - \frac{t'}{\pi} \right] \nonumber \\
& \equiv & \int_0^{\pi} \frac{dt}{\sqrt{t}} e^{-(\pi\eta)^2/t}
\left[ \theta_4(e^{-t\Delta}) \theta_4(e^{-t/\Delta}) - 1 \right] ,
\end{eqnarray}
where the Jacobi theta function $\theta_4(q) = \sum_j (-1)^j q^{j^2}$.
The asymptotic behaviors
\begin{equation}
\theta_4(e^{-t}) \mathop{\sim}_{t\to 0} 
\sqrt{\frac{\pi}{t}} e^{-\pi^2/(4t)} + \cdots ,
\qquad \theta_4(e^{-t}) \mathop{\sim}_{t\to\infty} 1 - 2 e^{-t} + \cdots
\end{equation}
ensure the convergence of the resulting integral.
To summarize this paragraph, the total energy (\ref{E}) can be rewritten 
as an integral over the finite interval $[0,\pi]$ as follows
\begin{eqnarray}
\frac{E(\Delta,\eta)}{e^2\sqrt{n}} & = & \frac{1}{2^{3/2}\sqrt{\pi}}
\int_0^{\pi} \frac{dt}{\sqrt{t}} \Bigg\{ 
2 \left[ \theta_3(e^{-t\Delta}) \theta_3(e^{-t/\Delta}) - 1 - \frac{\pi}{t} \right]
\nonumber \\ & &
+ e^{-\eta^2 t} \left[ \theta_2(e^{-t\Delta}) \theta_2(e^{-t/\Delta}) 
- \frac{\pi}{t} \right] + e^{-(\pi\eta)^2/t} 
\left[ \theta_4(e^{-t\Delta}) \theta_4(e^{-t/\Delta}) - 1 \right] \Bigg\} .
\label{EE}
\end{eqnarray}

The exact cancellation of singular terms near $t=0$ in this expression for
$E$ represents a numerical obstacle that should be circumvented. 
To accomplish the cancellation analytically, we shall consider the series 
representations of Jacobi theta functions and apply to them the Poisson 
transformation formula (\ref{Poisson});
after subtracting explicitly the singular term, the result appears as a
series of special functions.
In particular, for the first term in the integral (\ref{EE}) we obtain
\begin{eqnarray}
\int_0^{\pi} \frac{dt}{\sqrt{t}}
\left[ \theta_3(e^{-t\Delta}) \theta_3(e^{-t/\Delta}) - 1 - \frac{\pi}{t} \right]
& = & \int_0^{\pi} \frac{dt}{\sqrt{t}} \left[
\sum_{j,k} e^{-t j^2\Delta} e^{-t k^2/\Delta} - \frac{\pi}{t} \right] - 2\sqrt{\pi}
\nonumber \\ & = & \int_0^{\pi} dt \frac{\pi}{t^{3/2}} \left[
\sum_{j,k} e^{-(\pi j)^2/(\Delta t)} e^{-(\pi k)^2\Delta/t} - 1 \right] - 2\sqrt{\pi} .
\label{rep}
\end{eqnarray} 
The subtraction of the singularity is equivalent to the omission of 
the term $(j,k)=(0,0)$ from the summation.
Using the substitution $t'=t/\pi^2$ and introducing the function
\begin{equation} \label{z}
z_{\nu}(x,y) = \int_0^{1/\pi} \frac{dt}{t^{\nu}} e^{-x t} e^{-y/t} \qquad
\mbox{for $y>0$,}
\end{equation}
the last expression in Eq. (\ref{rep}) can be written as
\begin{equation}
2 \sum_{j=1}^{\infty} \left[ z_{3/2}(0,j^2/\Delta) + z_{3/2}(0,j^2\Delta) \right]
+ 4 \sum_{j,k=1}^{\infty} z_{3/2}(0,j^2/\Delta+k^2\Delta) - 2 \sqrt{\pi} .
\end{equation} 
Finally, performing the above procedure for all terms under integration 
in (\ref{EE}), we end up with the series representation
\begin{eqnarray}
\frac{E(\Delta,\eta)}{e^2\sqrt{n}} & = & \frac{1}{2^{3/2}\sqrt{\pi}}
\Bigg\{ 4\sum_{j=1}^{\infty} 
\left[ z_{3/2}(0,j^2/\Delta) + z_{3/2}(0,j^2\Delta) \right]
+ 8 \sum_{j,k=1}^{\infty} z_{3/2}(0,j^2/\Delta+k^2\Delta) \nonumber \\ & &
+ 2 \sum_{j=1}^{\infty} (-1)^j 
\left[ z_{3/2}((\pi\eta)^2,j^2/\Delta) + z_{3/2}((\pi\eta)^2,j^2\Delta) \right]
+ 4 \sum_{j,k=1}^{\infty} (-1)^j (-1)^k z_{3/2}((\pi\eta)^2,j^2/\Delta+k^2\Delta) 
\nonumber \\ & &
+ 4 \sum_{j,k=1}^{\infty} z_{3/2}(0,\eta^2+(j-1/2)^2/\Delta+(k-1/2)^2\Delta)
- 4 \sqrt{\pi} - \pi z_{1/2}(0,\eta^2) \Bigg\} . \label{series1}
\end{eqnarray}
The function $z_{\nu}(x,y)$ with $x=0$ is related to the so-called
Misra function\cite{Misra40,Born40} which was extensively used in 
single-layer lattice summations.\cite{Borwein88,Bowick06}
For our bilayer system with positive $\eta$ we need the more general 
function (\ref{z}) with $x\ge 0$. The convergence properties
of our series can be anticipated from the asymptotic relation
$z_\nu(0,y) \sim e^{-\pi y} \pi^{\nu-2}/y$

In numerical calculations, for a given $\eta$ we have to find such 
$\Delta^*$ which provides the minimum value of the energy (\ref{series1}). 
In practice, the series (\ref{series1}) must be cut at some $j,k=M$.
We document excellent convergence properties of the series (\ref{series1}) 
by considering the single-layer case $(\Delta=\sqrt{3},\eta=0)$ for which
the exact\cite{rque50}
\begin{equation}
E(\sqrt{3},0)/(e^2\sqrt{n}) = - 1.96051578931989165\ldots
\label{eq:E0}
\end{equation}
The cut at $M=1,2,3,4$ reproduces this exact value up to $2,5,10,17$ 
decimal digits, respectively.   
A similar accuracy is reached in all considered cases.
To be extremely accurate, we apply the $M=5$ cut everywhere,
and use the {\it Mathematica} software.

Another advantage of the series representation (\ref{series1}) is the
possibility of an explicit expansion of the function 
$E(\Delta^*,\eta)/(e^2\sqrt{n})$ around the controversial point $\eta=0$ 
and around the critical point $\eta_2^c$. 
As will be shown, this requires an analogous Taylor expansion of our 
$z$-functions.

\subsection{Going from phase I to phase II}
We know\cite{Bonsall,Jagla} that for the single layer, i.e. $\eta=0$, the structure providing
the minimum of the energy is the hexagonal lattice with $\Delta=\sqrt{3}$.
In what follows, we shall investigate the minimum of the energy (\ref{series1})
in the neighborhood of the point $(\Delta=\sqrt{3},\eta=0)$.
We set $\Delta=\sqrt{3}-\epsilon$ and consider $\epsilon$ to be 
infinitesimally small.
To derive a small-$\epsilon$ expansion of the energy (\ref{series1}),
we first perform this task for its series components.
From the integral definition (\ref{z}) it is easy to show that 
the $z$-functions under consideration exhibit an analytic (Taylor) 
expansion in $\epsilon$ of the form
\begin{equation}
z_{3/2}(0,j^2\Delta) = z_{3/2}(0,j^2\sqrt{3}) 
+ \epsilon j^2 z_{5/2}(0,j^2\sqrt{3}) 
+ \frac{1}{2} \epsilon^2 j^4 z_{7/2}(0,j^2\sqrt{3}) + {\cal O}(\epsilon^3) , 
\end{equation}
\begin{equation}
z_{3/2}(0,j^2/\Delta) = z_{3/2}(0,j^2/\sqrt{3}) 
- \epsilon \frac{j^2}{3} z_{5/2}(0,j^2/\sqrt{3}) 
+ \epsilon^2 \left[ \frac{j^4}{18} z_{7/2}(0,j^2/\sqrt{3}) -
\frac{j^2}{3^{3/2}} z_{5/2}(0,j^2/\sqrt{3}) \right] + {\cal O}(\epsilon^3) , 
\end{equation}
\begin{eqnarray}
z_{3/2}(0,j^2/\Delta+k^2\Delta) & = & z_{3/2}(0,j^2/\sqrt{3}+k^3\sqrt{3}) 
+ \epsilon \left( k^2 - \frac{j^2}{3} \right) z_{5/2}(0,j^2/\sqrt{3}+k^2\sqrt{3})
\nonumber \\ & & + \epsilon^2 \left[ \frac{1}{2} 
\left( k^2 - \frac{j^2}{3} \right)^2 z_{7/2}(0,j^2/\sqrt{3}+k^2\sqrt{3}) -
\frac{j^2}{3^{3/2}} z_{5/2}(0,j^2/\sqrt{3}+k^2\sqrt{3}) \right] 
+ {\cal O}(\epsilon^3) . 
\end{eqnarray}
Similar expansions can be derived for $z_{3/2}((\pi\eta)^2,j^2\Delta)$,
$z_{3/2}((\pi\eta)^2,j^2/\Delta)$, etc.
Inserting these expansions into (\ref{series1}), we obtain
\begin{equation} \label{epsilon1}
\frac{E(\sqrt{3}-\epsilon,\eta)}{e^2\sqrt{n}} 
= \frac{E(\sqrt{3},\eta)}{e^2\sqrt{n}}
+ f_1(\eta) \epsilon + f_2(\eta) \epsilon^2 + {\cal O}(\epsilon^3) ,
\end{equation}
where the explicit form of the prefactor functions $f_1(\eta)$ 
and $f_2(\eta)$ is written in the Appendix.

\begin{figure} [tb]
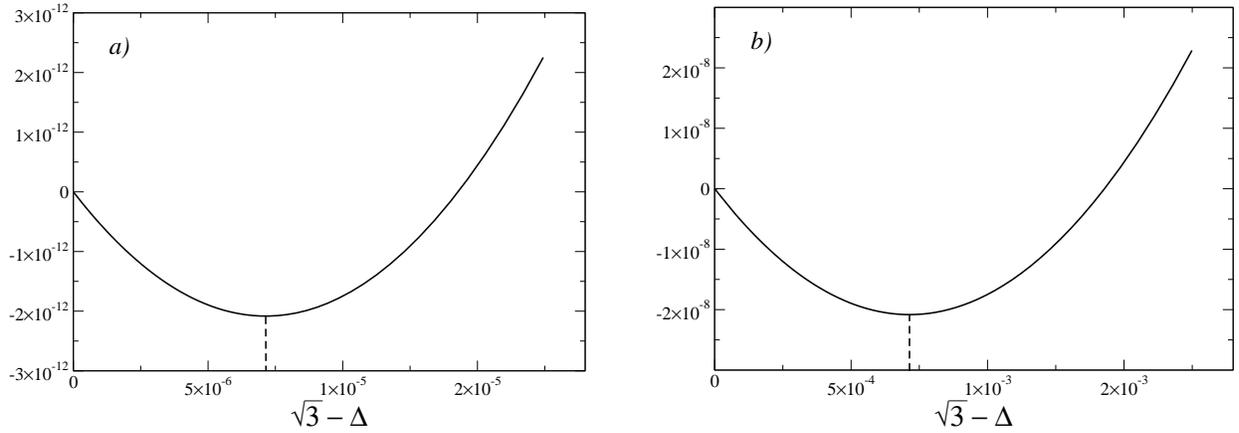

\includegraphics[width=0.43\textwidth]{smalleta3.eps} 
\hfil
\includegraphics[width=0.43\textwidth]{smalleta2.eps} 
\caption{The difference between the dimensionless energies 
$[E(\Delta,\eta)-E(\sqrt{3},\eta)]/(e^2\sqrt{n})$ versus 
$\epsilon=\sqrt{3}-\Delta$, calculated numerically by using (\ref{series1}) 
for two small values of $\eta$: a) $\eta=10^{-3}$ and b) $\eta=10^{-2}$.
For the range of aspect ratios chosen, these curves are indistinguishable 
from the analytical prediction $-0.5833 \, \eta^2 \epsilon+ 0.0408\, \epsilon^2$,
stemming from Eqs. (\ref{epsilon1}) and (\ref{eq:29}).
The values of $\epsilon^*$, which provide the energy minimum in the asymptotic
limit $\eta\to 0$ according to (\ref{transitionItoII}), are depicted by 
the vertical dashed lines for comparison.}
\label{fig:smalleta}
\end{figure}

Being close to the point $(\Delta=\sqrt{3},\eta=0)$, we are interested in 
the small-$\eta$ behavior of the  functions
$f_1(\eta)$ and $f_2(\eta)$.
The corresponding Taylor expansions in powers of $\eta^2$ can be performed 
explicitly, too.
The explicit form of $E(\sqrt{3},\eta)$ is here immaterial.
We find that $f_1(0)=0$, and, 
with a high precision, 
\begin{equation}
f_1(\eta) = - 0.5833059875\ldots\eta^2 + {\cal O}(\eta^4) , \qquad
f_2(\eta) =  0.0408440789\ldots + {\cal O}(\eta^2) .
\label{eq:29}
\end{equation}
For a fixed $\eta$, the extremum of the energy (\ref{epsilon1}) appears
at $\epsilon^*=\sqrt{3}-\Delta^*$ given by the stationarity condition 
\begin{equation}
\frac{\partial}{\partial\epsilon} \frac{E(\sqrt{3}-\epsilon,\eta)}{e^2\sqrt{n}}
\Bigg\vert_{\epsilon=\epsilon^*} = 0 = 
f_1(\eta) + 2 f_2(\eta) \epsilon^* .
\end{equation}
Namely,
\begin{equation} \label{transitionItoII}
\sqrt{3} - \Delta^* \equiv \epsilon^*(\eta) = 
- \frac{f_1(\eta)}{2 f_2(\eta)} = 7.14064\ldots \eta^2 + {\cal O}(\eta^4) .
\end{equation}
Since $\partial^2_{\epsilon}E(\sqrt{3}-\epsilon,\eta)\big\vert_{\epsilon=\epsilon^*} 
= f_2(\eta)>0$, the extremum is the minimum.
The result (\ref{transitionItoII}) tells us that howsoever small 
the dimensionless distance $\eta$ is, the buckled structure II with 
$\Delta<\sqrt{3}$ takes place.  
In other words, the structure I exists only strictly at $\eta_1^c=0$.
The fact that in the previous works\cite{Goldoni96,Weis01} the structure I 
was detected also for very small positive values of $\eta$ is probably related 
to extremely small values of the deviation $\epsilon^*\propto \eta^2$ 
for these $\eta$'s, which are ``invisible'' by standard numerical methods.
Like for instance, the structure I border reported
in\cite{Goldoni96} $\eta_1^c=0.006$ corresponds to 
$\epsilon^*=0.00026\ldots$.

In Fig. \ref{fig:smalleta}, we present the plots of the difference between 
the dimensionless energies $[E(\Delta,\eta)-E(\sqrt{3},\eta)]/(e^2\sqrt{n})$ 
versus $\epsilon=\sqrt{3}-\Delta$, calculated numerically by using 
(\ref{series1}) for two very small values of $\eta$: a) $\eta=10^{-3}$ 
and b) $\eta=10^{-2}$ which are well below/above the previous estimate 
of the phase I threshold,\cite{Goldoni96} respectively.
Alternatively, using the analytical expressions (\ref{epsilon1}) and (\ref{eq:29})
leads to the very same data.
The nonzero values of $\epsilon^*$, which provide the energy minima in 
the asymptotic limit $\eta\to 0$ according to formula (\ref{transitionItoII}), 
are depicted by the dashed lines for comparison.
We see that the energy minima fit well with the expected $\epsilon^*$
which is clear evidence for the phase I instability.
Note extremely small values $\propto 10^{-12}-10^{-8}$ of the energy difference, in Fig. \ref{fig:smalleta}, which justifies the derivation of an accurate formula for the
Coulombic energy. 

In Fig. \ref{fig:zoom12}, the asymptotic relation (\ref{transitionItoII}) 
(dashed line) is tested against numerical minimization of 
the energy (\ref{series1}) (solid curve) for small and intermediary 
values of $\eta$, in the logarithmic scale.
The agreement is very good, not only for small $\eta$, but in the
whole range of stability of phase II (it will be shown in the next 
subsection that phase II border is given by $\eta_2^c\simeq 0.26276\ldots$).

\subsection{Transition between phases II and III}
Going from phase I to phase II is not a phase transition in the usual sense.
However, the symmetry of the energy $E$ with respect to the transformation
$\Delta\to 1/\Delta$ has the fixed (self dual) point at $\Delta=1$ which is
the critical point of the phase transition from phase II to III.
Let us parameterize $\Delta$ as follows $\Delta=\exp(\epsilon)$.
The symmetry $\Delta\to 1/\Delta$ is now equivalent to 
$\epsilon\to -\epsilon$, i.e. the energy $E$ is an even function
of $\epsilon$.
The expansion of $E$ around the critical point $\Delta=1$ $(\epsilon=0)$
in small deviation $\epsilon$ follows from the representation (\ref{series1}):
\begin{equation} \label{phase23}
\frac{E(e^{\epsilon},\eta)}{e^2\sqrt{n}} =  \frac{E(1,\eta)}{e^2\sqrt{n}}
+ g_2(\eta) \epsilon^2 
+ g_4(\eta) \epsilon^4 + {\cal O}(\epsilon^6) .
\end{equation}
The explicit form of $g_2(\eta)$ is
written in the Appendix and $g_4(\eta)$ is not presented due to lack of space,
but has been derived.
The energy (\ref{phase23}) has the Ginzburg-Landau form, $\epsilon$ 
being the order parameter. 
In contrast to that mean field theory, the expression for our energy is exact.

\begin{figure} [t]
\begin{minipage}{80mm}
\includegraphics[width=1.00\textwidth]{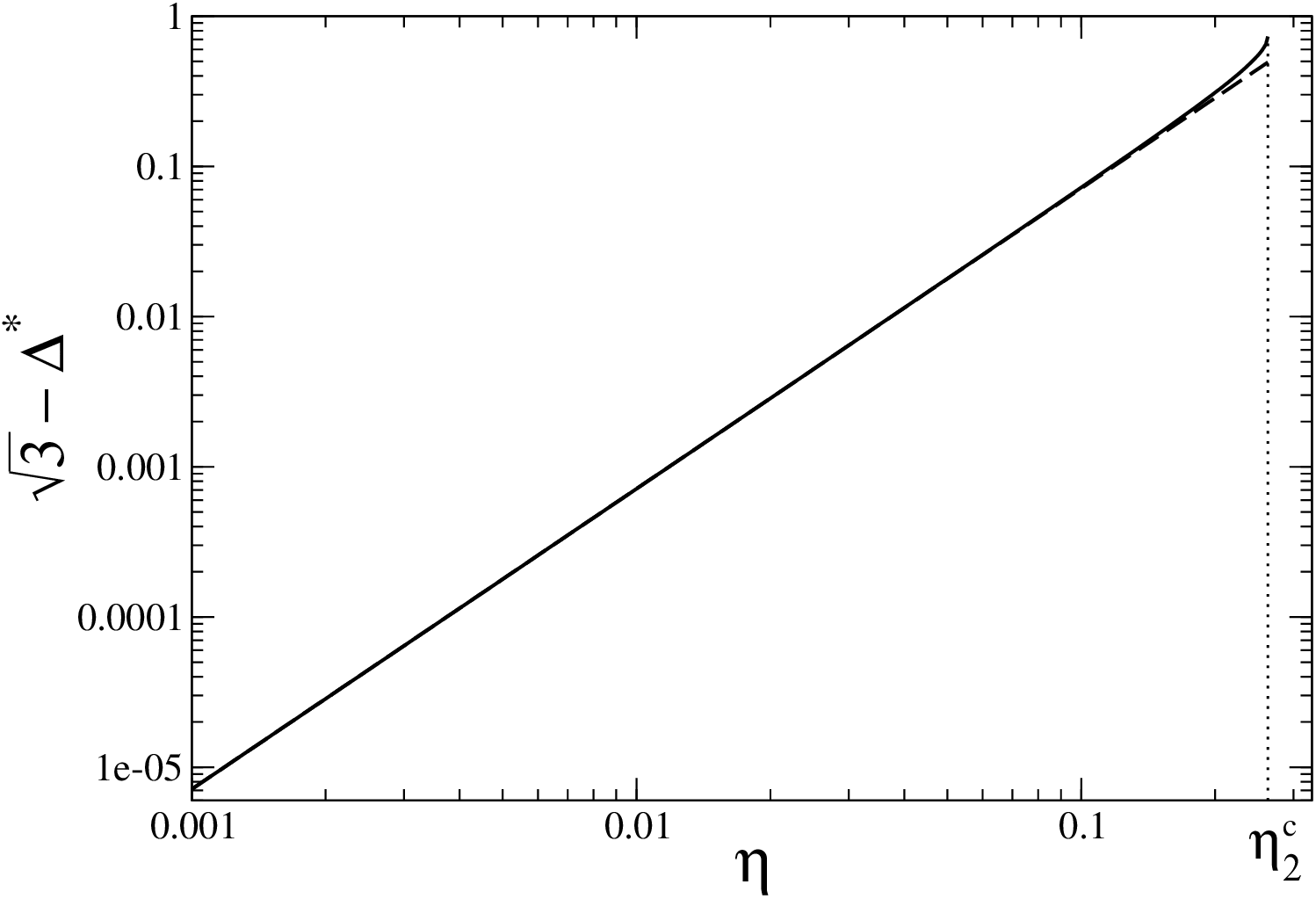}
\caption{Going from phase I to II: The test of the asymptotic relation 
(\ref{transitionItoII}) (dashed line) against numerical minimization of 
the energy (\ref{series1}) (solid curve) for small and intermediary 
values of $\eta$, in the logarithmic scale.}
\label{fig:zoom12}
\end{minipage}
\hfil
\begin{minipage}{80mm}
\includegraphics[width=1.00\textwidth]{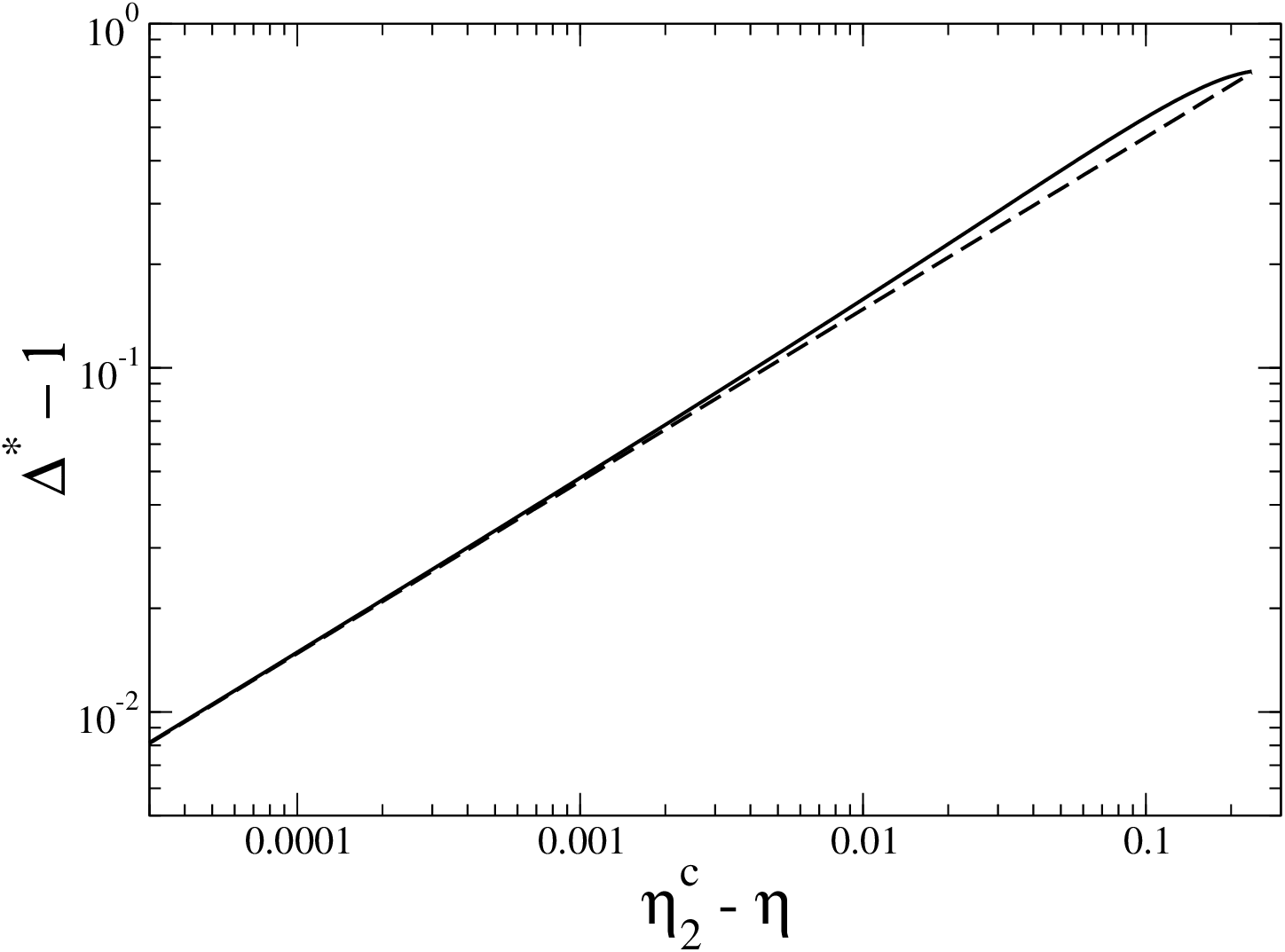}
\caption{Transition between phases II and III: The test of 
the asymptotic relation (\ref{transitionIItoIII}) (dashed line) against 
numerical minimization of the energy (\ref{series1}) (solid curve), 
in the logarithmic scale.}
\label{fig:zoom23}
\end{minipage}
\end{figure}

The critical point is associated with the vanishing of the prefactor
to $\epsilon^2$,
\begin{equation}
g_2(\eta)\Big\vert_{\eta=\eta_2^c} = 0 , \qquad \eta_2^c = 0.2627602682\ldots .
\end{equation}
The values of $\eta_2^c$ obtained in the previous studies were
$0.262$,\cite{Goldoni96}which is remarkably precise, $0.28$\cite{Weis01} and $0.27$.\cite{Lobaskin07} 
The functions $g_2(\eta)$ and $g_4(\eta)$ exhibit the following expansions
around the critical $\eta_2^c$:
\begin{equation}
g_2(\eta) = - 0.4620982808\ldots (\eta_2^c-\eta) + {\cal O}((\eta_2^c-\eta)^2) ,
\qquad g_4(\eta) = 0.1054378203\ldots + {\cal O}(\eta_2^c-\eta) .
\end{equation}
The extremum (minimum) of the energy (\ref{phase23}) appears at 
$\epsilon^*\simeq \Delta^* - 1$ given by the condition
\begin{equation} \label{eqn}
\frac{\partial}{\partial\epsilon} \frac{E(e^{\epsilon},\eta)}{e^2\sqrt{n}}
\Bigg\vert_{\epsilon=\epsilon^*} = 0 = 2 g_2(\eta) \epsilon^* 
+ 4 g_4(\eta) {\epsilon^*}^3 .
\end{equation}

For $\eta<\eta_2^c$ (the ``ordered'' phase II), we have one trivial
solution $\epsilon^*=0$ which however provides the local maximum
of the energy.
There exist two conjugate nontrivial solutions which yield the
needed energy minimum; considering one of these solutions, we arrive at
\begin{equation} \label{transitionIItoIII}
\Delta^* - 1 \simeq \epsilon^* = \left( 
- \frac{g_2(\eta)}{2 g_4(\eta)} \right)^{1/2} 
\simeq 1.48031 \sqrt{\eta_2^c-\eta} .
\end{equation}
The critical index $\beta$, describing the growth of the order parameter
from its zero critical value via $\epsilon^*\propto (\eta_2^c-\eta)^{\beta}$, 
has the mean field value $1/2$.
In Fig. \ref{fig:zoom23}, in the logarithmic scale, the asymptotic 
relation (\ref{transitionIItoIII}) (dashed line) is compared with 
the numerical minimization of the energy (\ref{series1}) (solid curve). 

For $\eta>\eta_2^c$ (the ``disordered'' phase III), we have the only
solution to (\ref{eqn}) $\epsilon^*=0$ (or equivalently $\Delta^*=1$) , 
i.e. the rigid phase III is stable, up to a transition to phase IV 
discussed in the next section.

\begin{figure} [t]
\begin{center}
\includegraphics[width=0.50\textwidth]{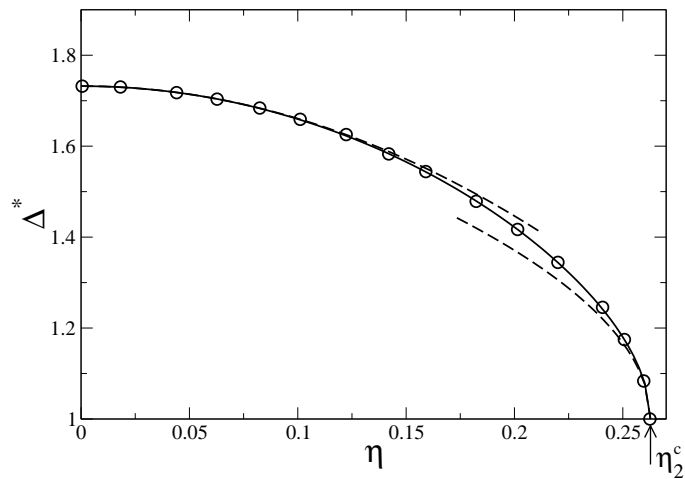}
\caption{The stability range of phase II: The plot of the lattice aspect 
ratio $\Delta^*$ versus $\eta$, obtained by the numerical minimization of 
the energy (\ref{series1}), is pictured by the solid curve.
Numerical data of Ref.\cite{Goldoni96} are presented 
by open circles.  
The asymptotic relations (\ref{transitionItoII}) for $\eta\to 0$ and 
(\ref{transitionIItoIII}) for $\eta\to \eta_2^c$ are depicted by dashed curves.}
\label{fig:Deltavseta}
\end{center}
\end{figure}

The plot of the lattice aspect ratio $\Delta^*$ versus $\eta$, obtained by 
the numerical minimization of the energy (\ref{series1}) in the whole 
stability range of the phase II, is pictured by the solid curve 
in Fig. \ref{fig:Deltavseta}.
$\Delta^*$ changes from $\sqrt{3}$ at $\eta=0$ to $1$ at $\eta=\eta_2^c$.
Numerical data of Goldoni and Peeters\cite{Goldoni96} (open circles)
are also presented for comparison.  
The asymptotic relations (\ref{transitionItoII}) for $\eta\to 0$ and 
(\ref{transitionIItoIII}) for $\eta\to \eta_2^c$ are also provided, for completeness.

\section{Phase IV} \label{sec:4}
In each of the two layers of phase IV (see Fig. \ref{fig:Structures}), 
the elementary cell is the rhombus with  angle $\varphi$ between 
the primitive translation vectors
\begin{equation}
\bm{a}_1 = a (1,0) , \qquad \bm{a}_2 = a (\cos\varphi,\sin\varphi) \qquad
{\rm with}\ a=\frac{1}{\sqrt{\sigma\sin\varphi}} .
\end{equation}
The lattice spacing $a$ is determined by the electroneutrality condition
$n_l=\sigma$; there is just one particle per rhombus of the surface
$a^2 \sin\varphi$ and so $n_l=1/(a^2\sin\varphi)$.
The special case of $\varphi=\pi/2$ corresponds to phase III.

\subsection{Energy of phase IV}
As before, the energy per particle $E$ of the bilayer structure consists 
of the intralayer and interlayer contributions, $E=E_{\rm intra}+E_{\rm inter}$.
As concerns the intralayer part, the 2D lattice vectors on one layer are 
indexed with respect to a reference particle on the same layer by 
${\bf r}(j,k) = j \bm{a}_1 + k \bm{a}_2$, where $j, k$ are any two integers
except for $(0,0)$.
The square of the lattice vector can be written as
\begin{equation}
\vert {\bf r}(j,k)\vert^2 = a^2 \left( j^2 + k^2 + 2 j k \cos\varphi \right)
= a^2 \left[ (j+k)^2 \cos^2(\varphi/2) + (j-k)^2 \sin^2(\varphi/2) \right] .
\end{equation}
This formula represents a kind of ``diagonalization'' of 
$\vert {\bf r}(j,k)\vert^2$ in indices. 
If $j+k$ is an even integer, we introduce new indices $n=(j+k)/2$ and 
$m=(j-k)/2$ covering all integers except for $(n,m)\ne (0,0)$.
If $j+k$ is an odd integer, we introduce indices $n=(j+k+1)/2$ and 
$n=(j-k+1)/2$ covering all integers.
Thus the interaction energy due to the Wigner crystal can be expressed as 
\begin{eqnarray} 
\frac{e^2}{2a} \sum_{j,k\atop (j,k)\ne (0,0)}
\frac{1}{\vert {\bf r}(j,k)\vert} & = &
\frac{e^2}{4a} \left[ \sum_{n,m\atop (n,m)\ne (0,0)}
\frac{1}{\sqrt{n^2\cos^2(\varphi/2) + m^2\sin^2(\varphi/2)}} \right. 
\nonumber \\ & &  \left.  + \sum_{n,m} 
\frac{1}{\sqrt{(n-1/2)^2\cos^2(\varphi/2) + (m-1/2)^2\sin^2(\varphi/2)}} 
\right] .
\end{eqnarray} 
Adding to this expression the interaction with the neutralizing background 
and using the gamma identity (\ref{gamma}) in close analogy with 
the previous section, we obtain
\begin{equation} \label{4intra}
\frac{E_{\rm intra}}{e^2\sqrt{n}} = \frac{1}{4\sqrt{\pi}}
\int_0^{\infty} \frac{dt}{\sqrt{t}} \left\{
\left[ \theta_3(e^{-t\delta}) \theta_3(e^{-t/\delta}) - 1 - \frac{\pi}{t} \right]
+ \left[ \theta_2(e^{-t\delta}) \theta_2(e^{-t/\delta}) - \frac{\pi}{t} \right] 
\right\} ,
\end{equation}
where $\delta = \tan(\varphi/2)$.

The Wigner lattices on the opposite layers are shifted with respect to
one another by the vector $(\bm{a}_1+\bm{a}_2)/2$.
To determine the interlayer contribution to the energy, we first consider
the square of the vector between the reference particle on one layer
and the vertices of the Wigner crystal on the other layer at distance $d$:
\begin{eqnarray}
\vert {\bf r}(j,k)\vert^2 & = & a^2 \left[ (j-1/2)^2 + (k-1/2)^2 
+ 2 (j-1/2) (k-1/2) \cos\varphi \right] + d^2 \nonumber \\  & = & 
a^2 \left[ (j+k-1)^2 \cos^2(\varphi/2) + (j-k)^2 \sin^2(\varphi/2) 
+ (d/a)^2 \right] .
\end{eqnarray}
Thus the interaction energy with the Wigner crystal reads
\begin{eqnarray} 
\frac{e^2}{2a} \sum_{j,k} \frac{1}{\vert {\bf r}(j,k)\vert} & = &
\frac{e^2}{4a} \left[ \sum_{n,m}
\frac{1}{\sqrt{(n-1/2)^2\cos^2(\varphi/2) + m^2\sin^2(\varphi/2)+d^2/(2a)^2}} 
\right. \nonumber \\ & &  \left.  + \sum_{n,m} 
\frac{1}{\sqrt{n^2\cos^2(\varphi/2) + (m-1/2)^2\sin^2(\varphi/2)+d^2/(2a)^2}} 
\right] .
\end{eqnarray} 
Adding the background term and using the gamma identity, we find that
\begin{equation} \label{4inter}
\frac{E_{\rm inter}}{e^2\sqrt{n}} = \frac{1}{4\sqrt{\pi}}
\int_0^{\infty} \frac{dt}{\sqrt{t}} e^{-\eta^2 t/2} \left\{
\left[ \theta_3(e^{-t\delta}) \theta_2(e^{-t/\delta}) - \frac{\pi}{t} \right]
+ \left[ \theta_2(e^{-t\delta}) \theta_3(e^{-t/\delta}) - \frac{\pi}{t} \right] 
\right\} .
\end{equation}

The total energy per particle $E$ is the sum of (\ref{4intra})
and (\ref{4inter}).
Note the invariance of $E$ with respect to the transformation 
$\delta\to 1/\delta$. 
With respect to the definition of $\delta = \tan(\varphi/2)$, 
this symmetry is equivalent to the obvious one $\varphi\to \pi-\varphi$.
Following subsequently similar lines as in previous sections,
the integral range $[0,\infty]$ can be reduced to $[0,\pi]$ by using 
the Poisson formula (\ref{Poisson}),
\begin{eqnarray}
\frac{E(\delta,\eta)}{e^2\sqrt{n}} & = & \frac{1}{4\sqrt{\pi}}
\int_0^{\pi} \frac{dt}{\sqrt{t}} \Bigg\{
2 \left[ \theta_3(e^{-t\delta}) \theta_3(e^{-t/\delta}) - 1 - \frac{\pi}{t} \right]
+ \left[ \theta_2(e^{-t\delta}) \theta_2(e^{-t/\delta}) - \frac{\pi}{t} \right] 
+ \left[ \theta_4(e^{-t\delta}) \theta_4(e^{-t/\delta}) - 1 \right] 
\nonumber \\ & & + e^{-\eta^2 t/2} 
\left[ \theta_3(e^{-t\delta}) \theta_2(e^{-t/\delta}) - \frac{\pi}{t} \right]
+ e^{-(\pi\eta)^2/(2t)} 
\left[ \theta_3(e^{-t\delta}) \theta_4(e^{-t/\delta}) - 1 \right]
\nonumber \\ & & + e^{-\eta^2 t/2} 
\left[ \theta_2(e^{-t\delta}) \theta_3(e^{-t/\delta}) - \frac{\pi}{t} \right] 
+ e^{-(\pi\eta)^2/(2t)} 
\left[ \theta_4(e^{-t\delta}) \theta_3(e^{-t/\delta}) - 1 \right] \Bigg\} .
\end{eqnarray}
Applying again the Poisson transformation formula (\ref{Poisson}) to the
series representations of Jacobi theta functions, the singular $t\to 0$
terms are canceled explicitly and we end up with the representation
of the energy per particle $E$ in terms of $z$-functions defined in (\ref{z}): 
\begin{eqnarray} \label{energy4}
\frac{E(\delta,\eta)}{e^2\sqrt{n}} & = & \frac{1}{2\sqrt{\pi}}
\Bigg\{ \sum_{j=1}^{\infty} [2+(-1)^j] 
\left[ z_{3/2}(0,j^2/\delta) + z_{3/2}(0,j^2\delta) \right]
+ 2 \sum_{j,k=1}^{\infty} [2+(-1)^j (-1)^k] z_{3/2}(0,j^2/\delta+k^2\delta) 
\nonumber \\ & &
+ 2 \sum_{j,k=1}^{\infty} z_{3/2}(0,(j-1/2)^2/\delta+(k-1/2)^2\delta)
+ \sum_{j=1}^{\infty} [1+(-1)^j] \left[ 
z_{3/2}((\pi\eta)^2/2,j^2/\delta) + z_{3/2}((\pi\eta)^2/2,j^2\delta) \right]
\nonumber \\ & &
+ 2 \sum_{j,k=1}^{\infty} [(-1)^j+(-1)^k] z_{3/2}((\pi\eta)^2/2,j^2/\delta+k^2\delta)
\nonumber \\ & &
+ 2 \sum_{j,k=1}^{\infty} \left[ z_{3/2}(0,\eta^2/2+(j-1/2)^2/\delta+k^2\delta)
+ z_{3/2}(0,\eta^2/2+(j-1/2)^2\delta+k^2/\delta) \right]
\nonumber \\ & &
+ \sum_{j=1}^{\infty} \left[ z_{3/2}(0,\eta^2/2+(j-1/2)^2/\delta) 
+ z_{3/2}(0,\eta^2/2+(j-1/2)^2\delta) \right]
- 3 \sqrt{\pi} - \pi z_{1/2}(0,\eta^2/2) \Bigg\} . \label{series2}
\end{eqnarray}

\subsection{Transition between phases III and IV}
The symmetry of the energy $E$ with respect to the transformation
$\delta\to 1/\delta$ has the fixed point at $\delta=1$ which is
the critical point of the phase transition between the phases III and IV.
Parameterizing $\delta$ as $\delta\equiv \tan(\varphi/2) = \exp(-\epsilon)$, 
the symmetry takes form $\epsilon\to -\epsilon$ and the energy $E$ 
is an even function of $\epsilon$.
The expansion of $E$ around the critical point $\delta=1$ (equivalent to 
$\theta=\pi/2$ or $\epsilon=0$) in small $\epsilon$ follows from 
the representation (\ref{series2}):
\begin{equation} \label{phase34}
\frac{E(e^{-\epsilon},\eta)}{e^2\sqrt{n}} = \frac{E(1,\eta)}{e^2\sqrt{n}}
+ h_2(\eta) \epsilon^2 
+ h_4(\eta) \epsilon^4 + {\cal O}(\epsilon^6) .
\end{equation}
The explicit form of $h_2(\eta)$ is
presented in the Appendix; The expression for $h_4(\eta)$ is too lengthy 
to be given, but is at our disposal.

The critical point is associated with the vanishing of the prefactor
of $\epsilon^2$,
\begin{equation}
h_2(\eta)\Big\vert_{\eta=\eta_3^c} = 0 , \qquad \eta_3^c = 0.6214809246\ldots .
\end{equation}
The values of $\eta_3^c$ obtained in the previous studies are
$0.622$,\cite{Goldoni96} $0.59$\cite{Weis01} and $0.62$.\cite{Lobaskin07} 
The functions $h_2(\eta)$ and $h_4(\eta)$ exhibit the following expansions
around the critical $\eta_3^c$:
\begin{equation}
h_2(\eta) = - 0.2675826391\ldots (\eta-\eta_3^c) + {\cal O}((\eta-\eta_3^c)^2) ,
\qquad h_4(\eta) = 0.0863245072\ldots + {\cal O}(\eta-\eta_3^c) .
\end{equation}
The extremum (minimum) of the energy (\ref{phase34}) appears at 
$\epsilon^*\simeq \pi/2 - \varphi^*$ given by the condition
\begin{equation}
\frac{\partial}{\partial\epsilon} \frac{E(e^{-\epsilon},\eta)}{e^2\sqrt{n}}
\Bigg\vert_{\epsilon=\epsilon^*} = 0 = 2 h_2(\eta) \epsilon^* 
+ 4 h_4(\eta) {\epsilon^*}^3 .
\end{equation}

\begin{figure} [t]
\begin{minipage}{80mm}
\includegraphics[width=1.00\textwidth]{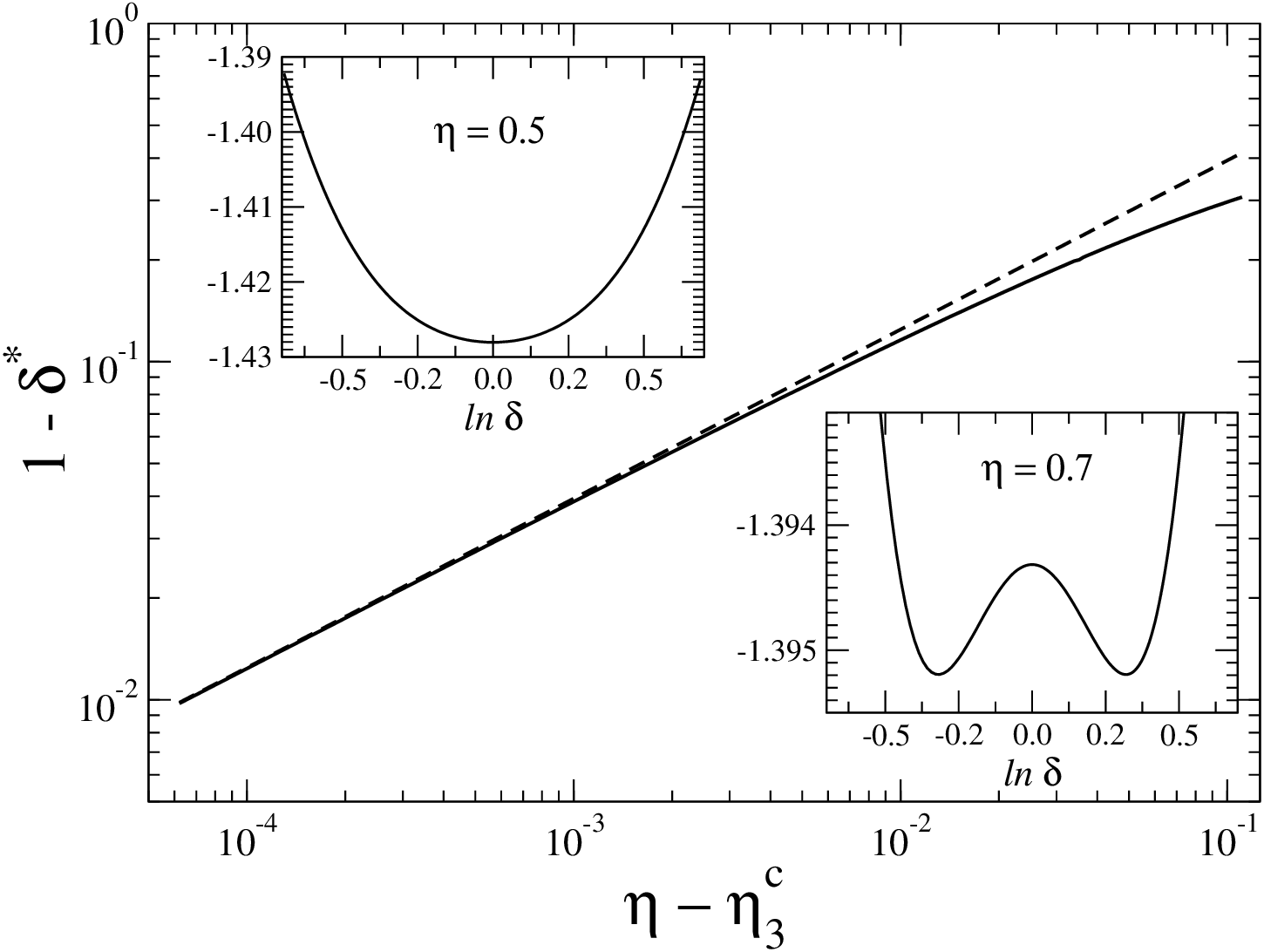}
\caption{Transition between phases III and IV: The test of 
the asymptotic relation (\ref{transitionIIItoIV}) (dashed line) against 
a numerical minimization of the energy (\ref{series2}) (solid curve), 
in the logarithmic scale.
The content of upper and lower insets is commented in the text.}
\label{fig:zoom34}
\end{minipage}
\hfil
\begin{minipage}{80mm}
\includegraphics[width=1.00\textwidth]{deltavseta.eps}
\caption{
Stability range of phase IV: The plot of the angle parameter $\delta^*$ 
versus $\eta$, obtained by the numerical minimization of the energy 
(\ref{series2}), is shown by the solid curve.
Numerical data of Ref.\cite{Goldoni96} are presented 
by open circles.  
The asymptotic relation (\ref{transitionIIItoIV}) for $\eta\to \eta_3^c$ 
is depicted by the dashed curve.}
\label{fig:deltavseta}
\end{minipage}
\end{figure}

For $\eta<\eta_3^c$ (``disordered'' phase III), we have the only
solution $\epsilon^*=0$ (or equivalently $\varphi^*=\pi/2$) which provides
the energy minimum, i.e. the rigid phase III is stable.
For $\eta>\eta_3^c$ (``ordered'' phase IV), the trivial solution 
$\epsilon^*=0$ becomes unstable.
The couple of conjugate nontrivial solutions, which provide 
the energy minimum, implies
\begin{equation} \label{transitionIIItoIV}
1-\delta^* \simeq \epsilon^* \simeq \frac{\pi}{2} - \varphi^* = \left( - 
\frac{h_2(\eta)}{2 h_4(\eta)} \right)^{1/2} \simeq 1.24494 \sqrt{\eta-\eta_3^c} .
\end{equation}
The critical index $\beta$ has again the mean field value $1/2$.
In Fig. \ref{fig:zoom34}, in a log-log scale,
the asymptotic relation (\ref{transitionIIItoIV}) (dashed line) is tested
against numerical minimization of the energy (\ref{series2}) (solid curve).
In the upper inset, we show the dependence of the energy on the logarithm 
of the angle parameter $\delta=\tan(\varphi/2)$ for $\eta=0.5$, where 
the phase III with $\delta=1$ is stable.
In the lower inset, the analogous plot is presented for $\eta=0.7$, 
where the phase IV with $\delta\ne 1$ is stable; phase III
corresponds in fact to a local maximum of the energy.
Note the symmetry of the energy with respect to the transformation
$\delta\to 1/\delta$ or, equivalently, $\ln\delta\to -\ln\delta$.

The plot of the angle parameter $\delta^*=\tan(\varphi^*/2)$ versus $\eta$, 
obtained by numerical minimization of the energy (\ref{series1}) in the whole 
stability range of the phase IV, is displayed by the solid curve 
in Fig. \ref{fig:deltavseta}.
$\delta^*$ changes from $1$ at $\eta=\eta_3^c$ (transition point from phase 
III to IV) to $\delta^c=0.69334\ldots$ at $\eta=\eta_4^c$ (transition point
from phase IV to V, see the next section).
Numerical data of Goldoni and Peeters\cite{Goldoni96} (open circles)
are presented for comparison.  
The asymptotic relation (\ref{transitionIIItoIV}) for $\eta\to \eta_3^c$ 
is depicted by the dashed curve.

\section{Phase V} \label{sec:5}
In a single layer of the phase V (see Fig. \ref{fig:Structures}), 
the elementary cell of the hexagonal lattice is the rhombus with 
the angle $\pi/3$ between the primitive translation vectors
\begin{equation}
\bm{a}_1 = a (1,0) , \qquad \bm{a}_2 = \frac{a}{2} (1,\sqrt{3}) \qquad
{\rm with}\ a=\frac{\sqrt{2}}{3^{1/4}} \frac{1}{\sqrt{\sigma}} .
\end{equation}
The lattice spacing $a$ follows from the electroneutrality condition
$n_l=\sigma$; there is just one particle per rhombus of surface
$\sqrt{3} a^2/2$, so that $n_l=2/(\sqrt{3} a^2)$.
Note that the images of vertices on the opposite layer are localized in 
the center of triangles and not rhombuses, as was the case of phase IV.
There is no continuous way to pass from phase IV to phase V.

\subsection{Energy of phase V}
To study the intralayer contribution to the energy of the reference particle
at the origin, we first consider the Wigner crystal of lattice vectors 
${\bf r}(j,k)=j \bm{a}_1 + k \bm{a}_2$, where integers $(j,k)\ne (0,0)$.
The square of the lattice vector is expressible as
\begin{equation}
\vert {\bf r}(j,k)\vert^2 = a^2 \left( j^2 + k^2 + j k \right)
= \frac{a^2}{4} \left[ 3 (j+k)^2 + (j-k)^2 \right] .
\end{equation}
In analogy with phase IV, we introduce new $n,m$ indices for each of 
the cases $j+k$ being an even and odd integer. 
The interaction energy due to the Wigner crystal then reads 
\begin{equation} 
\frac{e^2}{2a} \sum_{j,k\atop (j,k)\ne (0,0)} \frac{1}{\vert {\bf r}(j,k)\vert} 
= \frac{e^2}{2a} \left[ \sum_{n,m\atop (n,m)\ne (0,0)} \frac{1}{\sqrt{3 n^2+ m^2}}
+ \sum_{n,m} \frac{1}{\sqrt{3(n-1/2)^2 + (m-1/2)^2}} \right] .
\end{equation} 
Adding to this expression the interaction with the neutralizing background 
and using the gamma identity, we find
\begin{equation} \label{5intra}
\frac{E_{\rm intra}}{e^2\sqrt{n}} = \frac{1}{4\sqrt{\pi}}
\int_0^{\infty} \frac{dt}{\sqrt{t}} \left\{
\left[ \theta_3(e^{-t/\sqrt{3}}) \theta_3(e^{-t\sqrt{3}}) - 1 - \frac{\pi}{t} \right]
+ \left[ \theta_2(e^{-t/\sqrt{3}}) \theta_2(e^{-t\sqrt{3}}) - \frac{\pi}{t} \right] 
\right\} .
\end{equation}

The hexagonal lattices on the opposite layers are shifted with respect to
one another by the vector $(\bm{a}_1+\bm{a}_2)/3$; note that the factor $1/3$
differs from $1/2$ of the previous phases I-IV.
To determine the interlayer contribution to the energy, we first consider
the square of the vector between the reference particle on one layer
and the vertices of the Wigner crystal on the other layer at distance $d$:
\begin{equation}
\vert {\bf r}(j,k)\vert^2 = a^2 
\left[ (j+1/3)^2 + (k+1/3)^2 + (j+1/3) (k+1/3) \right] + d^2 =
\frac{a^2}{4} \left[ 3 (j+k+2/3)^2 + (j-k)^2 \right] + d^2 .
\end{equation}
Going from $(j,k)$ to integers $(n,m)$, the interaction energy with 
the Wigner crystal on the opposite side takes the form
\begin{equation} 
\frac{e^2}{2a} \sum_{j,k} \frac{1}{\vert {\bf r}(j,k)\vert} =
\frac{e^2}{2a} \left[ \sum_{n,m} \frac{1}{\sqrt{3 (n+1/3)^2 + m^2+(d/a)^2}} 
+ \sum_{n,m} \frac{1}{\sqrt{3 (n-1/6)^2 + (m-1/2)^2 +(d/a)^2}} \right] .
\end{equation} 
Adding the background term, using the gamma identity and the readily
derivable relations
\begin{equation}
\sum_j e^{-3t(j+1/3)^2} = \frac{1}{2} \left[
\theta_3(e^{-t/3}) - \theta_3(e^{-3t}) \right] , \qquad
\sum_j e^{-3t(j-1/6)^2} = \frac{1}{2} \left[
\theta_2(e^{-t/3}) - \theta_2(e^{-3t}) \right] ,
\end{equation} 
we find that
\begin{eqnarray}
\frac{E_{\rm inter}}{e^2\sqrt{n}} & = & \frac{1}{4\sqrt{\pi}}
\int_0^{\infty} \frac{dt}{\sqrt{t}} 
\left( -\frac{1}{2} e^{-\eta^2 t/2} + \frac{\sqrt{3}}{2} e^{-3\eta^2 t/2} \right)
\nonumber \\ & & \times \left\{ 
\left[ \theta_3(e^{-t/\sqrt{3}}) \theta_3(e^{-t\sqrt{3}}) - 1 - \frac{\pi}{t} \right]
+ \left[ \theta_2(e^{-t/\sqrt{3}}) \theta_2(e^{-t\sqrt{3}}) - \frac{\pi}{t} \right] 
\right\} . \label{5inter}
\end{eqnarray}

The total energy per particle $E$ is given by the sum of (\ref{5intra})
and (\ref{5inter}).
The Poisson formula (\ref{Poisson}) enables us to reduce the integral 
range to $[0,\pi]$, 
\begin{eqnarray}
\frac{E(\eta)}{e^2\sqrt{n}} & = & \frac{1}{4\sqrt{\pi}} 
\int_0^{\pi} \frac{dt}{\sqrt{t}} \Bigg\{ 
\left( 1 - \frac{1}{2} e^{-\eta^2 t/2} + \frac{\sqrt{3}}{2} e^{-3\eta^2 t/2} \right)
\left[ \theta_3(e^{-t/\sqrt{3}}) \theta_3(e^{-t\sqrt{3}}) - 1 - \frac{\pi}{t} \right]
\nonumber \\ & & +
\left( 1 - \frac{1}{2} e^{-(\pi\eta)^2/(2t)} 
+ \frac{\sqrt{3}}{2} e^{-3(\pi\eta)^2/(2t)} \right)
\left[ \theta_3(e^{-t/\sqrt{3}}) \theta_3(e^{-t\sqrt{3}}) - 1 - \frac{\pi}{t} \right]
\nonumber \\ & & + 
\left( 1 - \frac{1}{2} e^{-\eta^2 t/2} + \frac{\sqrt{3}}{2} e^{-3\eta^2 t/2} \right)
\left[ \theta_2(e^{-t/\sqrt{3}}) \theta_2(e^{-t\sqrt{3}}) - \frac{\pi}{t} \right] 
\nonumber \\ & & + 
\left( 1 - \frac{1}{2} e^{-(\pi\eta)^2/(2t)} 
+ \frac{\sqrt{3}}{2} e^{-3(\pi\eta)^2/(2t)} \right)
\left[ \theta_4(e^{-t/\sqrt{3}}) \theta_4(e^{-t\sqrt{3}}) - 1 \right] \Bigg\} .
\label{eq:59}
\end{eqnarray}
In terms of the functions
\begin{eqnarray}
I_2(x,y) & \equiv & \int_0^{\pi} \frac{dt}{\sqrt{t}} e^{-x t/\pi^2} e^{-y \pi^2/t} 
\left[ \theta_2(e^{-t/\sqrt{3}}) \theta_2(e^{-t\sqrt{3}}) - \frac{\pi}{t} \right] 
\nonumber \\ & = & 2 \sum_{j=1}^{\infty} (-1)^j 
\left[ z_{3/2}(x,y+j^2/\sqrt{3}) + z_{3/2}(x,y+j^2\sqrt{3}) \right]
+ 4 \sum_{j,k=1}^{\infty} (-1)^j (-1)^k z_{3/2}(x,y+j^2/\sqrt{3}+k^2\sqrt{3}) , 
\phantom{a}
\end{eqnarray}
\begin{eqnarray}
I_3(x,y) & \equiv & \int_0^{\pi} \frac{dt}{\sqrt{t}} e^{-x t/\pi^2} e^{-y \pi^2/t} 
\left[ \theta_3(e^{-t/\sqrt{3}}) \theta_3(e^{-t\sqrt{3}}) -1-\frac{\pi}{t} \right] 
\nonumber \\ & = & 2 \sum_{j=1}^{\infty}
\left[ z_{3/2}(x,y+j^2/\sqrt{3}) + z_{3/2}(x,y+j^2\sqrt{3}) \right]
+ 4 \sum_{j,k=1}^{\infty} z_{3/2}(x,y+j^2/\sqrt{3}+k^2\sqrt{3}) - \pi z_{1/2}(x,y) , 
\end{eqnarray}
\begin{eqnarray}
I_4(x,y) & \equiv & \int_0^{\pi} \frac{dt}{\sqrt{t}} e^{-x t/\pi^2} e^{-y \pi^2/t} 
\left[ \theta_4(e^{-t/\sqrt{3}}) \theta_4(e^{-t\sqrt{3}}) -1 \right] 
\nonumber \\ & = & 4 \sum_{j,k=1}^{\infty} 
z_{3/2}(x,y+(j-1/2)^2/\sqrt{3}+(k-1/2)^2\sqrt{3}) - \pi z_{1/2}(x,y) ,
\end{eqnarray}
the energy per particle $E$ is expressible as
\begin{eqnarray} \label{energy5}
\frac{E(\eta)}{e^2\sqrt{n}} & = & \frac{1}{4\sqrt{\pi}} \Bigg\{ \left[ 
2 I_3(0,0) - \frac{1}{2} I_3((\pi\eta)^2/2,0) - \frac{1}{2} I_3(0,\eta^2/2)
+ \frac{\sqrt{3}}{2} I_3(3(\pi\eta)^2/2,0) 
+ \frac{\sqrt{3}}{2} I_3(0,3\eta^2/2) \right] \nonumber \\ & &
+ \left[ I_2(0,0) - \frac{1}{2} I_2((\pi\eta)^2/2,0) 
+ \frac{\sqrt{3}}{2} I_2(3(\pi\eta)^2/2,0) \right]
+ \left[ I_4(0,0) - \frac{1}{2} I_4(0,\eta^2/2) 
+ \frac{\sqrt{3}}{2} I_4(0,3\eta^2/2) \right] \Bigg\} .
\end{eqnarray}

\subsection{Transition between phases IV and V}
Increasing $\eta$ from $\eta_3^c$, phase IV is stable up to the point
$\eta_4^c$ at which the energy of phase IV (\ref{energy4}), evaluated
at $\delta^*$ which minimizes this energy, equals to the energy of
phase V (\ref{energy5}). 
Our result is
\begin{equation}
\eta_4^c = 0.73242\ldots .
\end{equation}
The values of $\eta_4^c$ obtained in the previous studies were relatively dispersed:
$0.732$,\cite{Goldoni96} $0.70$\cite{Weis01} and $0.87$.\cite{Lobaskin07}

The phase transition is of first order since the energies of phases IV
and V have as functions of $\eta$ different slopes which causes
the discontinuity of the first derivative of the energy with respect to 
$\eta$ at the transition point $\eta_4^c$.
The angle parameter $\delta$, which minimizes the energy of the phase IV 
at the critical point $\eta_4^c$, is found to be $\delta^c=0.69334\ldots$.
Since $\delta = \tan(\varphi/2)$, the corresponding angle
$\varphi^c = 69.4702\ldots^{\circ}$; this angle is very close to the
estimate $\varphi^c = 69.48^{\circ}$ of the work.\cite{Goldoni96}
Going from phase IV to V, the angle skips to $60^{\circ}$ as is indicated
in Fig. \ref{fig:deltavseta}.  

\subsection{Discussion}
Two different ``sum-rules'' can be derived, that allow for a critical assessment
of the results obtained. The simplest one relies on the geometrical
proximity between structures I (a single hexagonal crystal) and V
(two hexagonal crystals at half density). For large distances, the two
crystals decouple and we have, making use of straightforward notations, 
\begin{equation}
 E_{I}(\sqrt{3},\eta=0) = \sqrt{2} \, E_{V}(\eta \to \infty).
\end{equation}
With our series representations (\ref{series1}) for  $E_{I}$
and (\ref{energy5}) for $E_{V}$, this identity holds.
Another more subtle constraint follows from a combination of elementary geometric considerations,\cite{Rouzina}
which impose that 
\begin{equation}
 E_{V}(\eta=0) \,=\, \frac{1+\sqrt{3}}{2\sqrt{2}} \, E_{I}\left(\sqrt{3}, \eta=0\right).
\end{equation}
We have also checked that this identity holds with the expressions provided above
(note though that $\eta=0$ lies outside the stability range of structure V).

Finally, it is interesting to consider both the large small and 
large distance behavior 
of the energy. For small $\eta$, it can be shown that both 
structures I and II share the same energy expansion,
up to order $\eta^3$ included:
\begin{equation}
E_{II}(\Delta^*,\eta) = E_I(\eta) + {\cal O}(\eta^4)
\end{equation}
where $\Delta^*$ is the previously introduced optimal 
aspect ratio that minimizes $E_{II}(\Delta^*,\eta)$ for a 
given $\eta$. Explicit calculation up to order $\eta^2$
shows that 
\begin{equation}
\frac{E_{II}(\Delta^*,\eta)}{e^2 \sqrt{n}} \, = \, 
\frac{E_{II}(\sqrt{3},0)}{e^2 \sqrt{n}} \,+\,
\frac{\pi \,\eta}{\sqrt{2}} \,- 2.59372\ldots \, \eta^2 +{\cal O}(\eta^3),
\label{eq:smalletaexp}
\end{equation}
where the precise value of $E_{II}(\sqrt{3},0)=E_I(0)$ has been given in Eq. 
(\ref{eq:E0}). We note that the linear term in (\ref{eq:smalletaexp})
generates a contact pressure $-2\pi \sigma^2 e^2$
for $\eta\to 0$. A similar term was reported in,\cite{Goldoni96} 
where however the term in $\eta^2$ differs from ours by a large factor
(0.2122 instead of 2.5937)

At large distances, the relevant phase is structure V, from which
the inter-plate pressure follows. The large $\eta$ case 
is encoded in the small $t$ limit of Eq. (\ref{eq:59}), or equivalently, 
Eq. (\ref{5inter}). A saddle point argument leads to
\begin{equation}
 \frac{E_{V}(\eta)}{e^2 \sqrt{n}} \, \sim \, \frac{E_{V}(\infty)}{e^2 \sqrt{n}} \,
 -  \, \frac{3^{5/4}}{4} \, \exp\left( -\frac{4 \pi}{\sqrt{2}\, 3^{1/4}} \, \eta
  \right).
\end{equation}
We recover an expression already obtained in,\cite{Goldoni96} 
at variance with other
approaches.\cite{EsKa95} Taking the $\eta$ derivative 
and remembering that $n=2\sigma$
yields the inter-plate pressure
\begin{equation}
P \, = \,  -2\,\sigma \frac{\partial E_{V}}{\partial d} 
\,=\, -2\sigma^{3/2} \frac{\partial E_{V}}{\partial \eta}
\,\sim\, -6 \pi (\sigma e)^2 \, 
\exp
\left(-\frac{4 \pi}{\sqrt 2 \, 3^{1/4}} \, \eta\right).
\end{equation}
The $\eta$ dependence is well known, since it can be written $\exp(-G_0 d)$, 
with $G_0$ the modulus of the first reciprocal lattice vector. 
It should be noted though that the prefactor differs from the often 
reported $2\pi (\sigma e)^2$ (see e.g. \cite{LaLP00}), by a 
factor 3.

\section{Conclusion} \label{sec:conclusion}
The system of classical charged particles, forming a sequence of 
bilayer Wigner structures at zero temperature as the distance between
the plates is increasing, has a rather long history.
We have presented here a new method to calculate the Coulomb ground-state 
energy of each Wigner structure.
Based on a series of transformations and using general properties of
the Jacobi theta functions, we expressed the energies in terms of
quickly converging series of the functions $z_{\nu}(x,y)$ defined in (\ref{z}).
The presence of the neutralizing background manifests itself simply as
the subtraction of singularities of the Jacobi theta functions under
an auxiliary integration.

\begin{figure} [htb]
\includegraphics[width=0.52\textwidth]{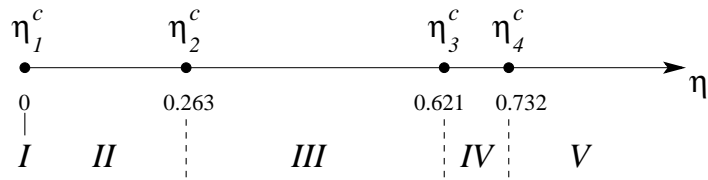} 
\caption{Summary of phase transition scenario. The rounded off values of the
thresholds are mentioned. Structure I is realized at $\eta=0$ only
(vanishing inter-plate separation).}
\label{fig:summary}
\end{figure}

Numerical evaluation of the series requires modest computer and programming 
facilities, and at the same time provides extremely accurate estimates of 
the energy. We took advantage of this feature, supplemented by analytical
work, to improve and complete previous studies 
in three aspects:
\begin{itemize}
\item
There was a relatively large dispersion in the determination of 
the transition points between phases; this concerns especially 
the first-order phase transition between phases IV and V. 
Our method improves significantly the location of all transition points,
that can be worked out with arbitrary precision.
Figure \ref{fig:summary} gives an overview of the sequence
of phases together with the corresponding thresholds.
\item
We resolved, analytically and numerically, a previous
controversy about the stability phase I, thereby corroborating
the findings of Ref.\cite{Messina03}
We found that this phase is stable only at zero distance between the
plates, $\eta=0=d$.
To confirm numerically this result, we worked with extremely small values 
of the energy differences $\propto 10^{-12}-10^{-8}$ for distances 
$\eta=10^{-3}$ and $10^{-2}$ (see Fig. \ref{fig:smalleta}), which are 
``invisible'' by standard numerical methods.
The agreement between the $\eta\to 0$ asymptotic relation
(\ref{transitionItoII}), calculated analytically by using 
the Taylor expansions of the functions $z_{\nu}(x,y)$, and 
the numerical minimization of the energy, presented in Fig. \ref{fig:zoom12},
is excellent.
\item
The expansions of the structure energies around second-order transition 
points can be done analytically which enables us to specify the critical 
phenomena at the phase transition points; see the expansions pertaining
to the transitions from phase II to III
(\ref{transitionIItoIII}) and from phase III to IV (\ref{transitionIIItoIV}).
The agreement between these analytic formulas and numerical minimization 
of the ground-state energy is very good; It can be appreciated in Figs. \ref{fig:zoom23} and 
\ref{fig:zoom34}. Quite expectedly for a zero temperature situation, 
the critical behavior is always of the Ginzburg-Landau type, 
with the mean-field critical index $\beta=1/2$ for the growth
of the order parameters in the ``ordered'' phases.   
\end{itemize}

It is clear that our method can be directly generalized to other problems 
concerning the lattice summations over pair interactions, not only
the Coulomb ones.
The bilayers with repulsive Yukawa interactions, extensively studied in 
the past,\cite{Messina03,Mazars11} or with inverse power laws,\cite{Mazars10} deserve our attention.
We additionally emphasize that the ground states under consideration here are such that the ions are distributed
evenly (50\% on each plate): in other words, the ionic surface density of 
one Wigner crystal on a given plate is $\sigma$, and coincides with the plate 
homogeneous surface 
density. When dealing with asymmetric plates, this local neutrality
assumption should be relaxed,\cite{Messina00,Messina09} 
still enforcing global neutrality.
This makes the asymmetric problem significantly more complex, 
and an interesting perspective
for future work. Finally, consideration of dielectric jumps between the walls 
and the interstitial slab is also a relevant venue for forthcoming
investigations.

\begin{acknowledgments}
We would like to thank M. Mazars and C. Texier for useful discussions.
L. \v{S}. is grateful to LPTMS for their kind hospitality.
The support received from the grants VEGA No. 2/0049/2012 and CE-SAS QUTE
is acknowledged.
\end{acknowledgments}

\renewcommand{\theequation}{A\arabic{equation}}
\setcounter{equation}{0}

\section*{APPENDIX}

The prefactor functions $f_1(\eta)$ and $f_2(\eta)$ of
the expansion (\ref{epsilon1}) read
\begin{eqnarray}
f_1(\eta) & = & \frac{1}{2^{3/2}\sqrt{\pi}} \Bigg\{ 4 \sum_{j=1}^{\infty} j^2 
\left[ z_{5/2}(0,j^2\sqrt{3})  - \frac{1}{3} z_{5/2}(0,j^2/\sqrt{3}) \right] 
+ 8 \sum_{j,k=1}^{\infty} \left( k^2 - \frac{j^3}{3} \right) 
z_{5/2}(0,j^2/\sqrt{3}+k^2\sqrt{3}) \nonumber \\ & &
+ 2 \sum_{j=1}^{\infty} (-1)^j j^2 \left[ z_{5/2}((\pi\eta)^2,j^2\sqrt{3}) 
- \frac{1}{3} z_{5/2}((\pi\eta)^2,j^2/\sqrt{3}) \right] \nonumber \\ & &
+ 4 \sum_{j,k=1}^{\infty} (-1)^j (-1)^k \left( k^2 - \frac{j^2}{3} \right) 
z_{5/2}((\pi\eta)^2,j^2/\sqrt{3}+k^2\sqrt{3}) \nonumber \\ & &
+ 4 \sum_{j,k=1}^{\infty} \left[ \left( k-\frac{1}{2} \right)^2 -
\frac{1}{3} \left( j-\frac{1}{2} \right)^2 \right]
z_{5/2}(0,\eta^2+(j-1/2)^2/\sqrt{3}+(k-1/2)^2\sqrt{3}) \Bigg\} ,
\end{eqnarray}
\begin{eqnarray}
f_2(\eta) & = & \frac{1}{2^{3/2}\sqrt{\pi}} \Bigg\{ 4 \sum_{j=1}^{\infty} 
\left[ \frac{j^4}{2} z_{7/2}(0,j^2\sqrt{3}) +
\frac{j^4}{18} z_{7/2}(0,j^2/\sqrt{3})  
- \frac{j^2}{3^{3/2}} z_{5/2}(0,j^2/\sqrt{3}) \right] \nonumber \\ & &
+ 8 \sum_{j,k=1}^{\infty} \left[ \frac{1}{2} \left( k^2 - \frac{j^3}{3} \right)^2 
z_{7/2}(0,j^2/\sqrt{3}+k^2\sqrt{3}) - \frac{j^2}{3^{3/2}} 
z_{5/2}(0,j^2/\sqrt{3}+k^2\sqrt{3}) \right] \nonumber \\ & &
+ 2 \sum_{j=1}^{\infty} (-1)^j \left[
\frac{j^4}{2} z_{7/2}((\pi\eta)^2,j^2\sqrt{3}) 
+ \frac{j^4}{18} z_{7/2}((\pi\eta)^2,j^2/\sqrt{3}) 
- \frac{j^2}{3^{3/2}} z_{5/2}((\pi\eta)^2,j^2/\sqrt{3}) \right] \nonumber \\ & &
+ 4 \sum_{j,k=1}^{\infty} (-1)^j (-1)^k \left[ 
\frac{1}{2} \left( k^2 - \frac{j^3}{3} \right)^2 
z_{7/2}((\pi\eta)^2,j^2/\sqrt{3}+k^2\sqrt{3}) 
- \frac{j^2}{3^{3/2}} z_{5/2}((\pi\eta)^2,j^2/\sqrt{3}+k^2\sqrt{3}) \right] 
\nonumber \\ & & + 4 \sum_{j,k=1}^{\infty} 
\Bigg[ \frac{1}{2} \left[ \left( k-\frac{1}{2} \right)^2 -
\frac{1}{3} \left( j-\frac{1}{2} \right)^2 \right]^2
z_{7/2}(0,\eta^2+(j-1/2)^2/\sqrt{3}+(k-1/2)^2\sqrt{3}) \nonumber \\ & &
- \frac{1}{3^{3/2}} \left( j - \frac{1}{2} \right)^2
z_{5/2}(0,\eta^2+(j-1/2)^2/\sqrt{3}+(k-1/2)^2\sqrt{3}) \Bigg] \Bigg\} .
\end{eqnarray}

The prefactor function $g_2(\eta)$ of the expansion (\ref{phase23}) takes the form
\begin{eqnarray}
g_2(\eta) & = & \frac{1}{\sqrt{2\pi}} \Bigg\{ 2 \sum_{j=1}^{\infty} 
\left[ j^4 z_{7/2}(0,j^2) - j^2 z_{5/2}(0,j^2) \right] \nonumber \\ & &
+ 2 \sum_{j,k=1}^{\infty} \left[ (j^2-k^2)^2 z_{7/2}(0,j^2+k^2) 
- (j^2+k^2) z_{5/2}(0,j^2+k^2) \right] \nonumber \\ & &
+ \sum_{j=1}^{\infty} (-1)^j \left[ j^4 z_{7/2}((\pi\eta)^2,j^2) 
- j^2 z_{5/2}((\pi\eta)^2,j^2) \right] \nonumber \\ & &
+ \sum_{j,k=1}^{\infty} (-1)^j (-1)^k 
\left[ (j^2-k^2)^2 z_{7/2}((\pi\eta)^2,j^2+k^2) 
- (j^2+k^2) z_{5/2}((\pi\eta)^2,j^2+k^2) \right] \nonumber \\ & &
+ \sum_{j,k=1}^{\infty} \left( \left[ (j-1/2)^2-(k-1/2)^2 \right]^2 
z_{7/2}(0,\eta^2+(j-1/2)^2+(k-1/2)^2) \right. \nonumber \\ & & \left.
- \left[ (j-1/2)^2+(k-1/2)^2 \right] 
z_{5/2}(0,\eta^2+(j-1/2)^2+(k-1/2)^2) \right) \Bigg\} .
\end{eqnarray}

Finally, the prefactor function $h_2(\eta)$ of the expansion (\ref{phase34})
can be written
\begin{eqnarray}
h_2(\eta) & = & \frac{1}{2\sqrt{\pi}} \Bigg\{
2 \sum_{j=1}^{\infty} \left[ j^4 z_{7/2}(0,j^2) - j^2 z_{5/2}(0,j^2) \right]
+ \sum_{j=1}^{\infty} (-1)^j \left[ j^4 z_{7/2}(0,j^2) - j^2 z_{5/2}(0,j^2) \right]
\nonumber \\ & &
+ 2 \sum_{j,k=1}^{\infty} \left[ (j^2-k^2)^2 z_{7/2}(0,j^2+k^2) 
- (j^2+k^2) z_{5/2}(0,j^2+k^2) \right] \nonumber \\ & &
+ \sum_{j,k=1}^{\infty} (-1)^j (-1)^k \left[ (j^2-k^2)^2 z_{7/2}(0,j^2+k^2) 
- (j^2+k^2) z_{5/2}(0,j^2+k^2) \right] \nonumber \\ & &
+ \sum_{j,k=1}^{\infty} \Big( \left[ (j-1/2)^2-(k-1/2)^2 \right]^2
z_{7/2}(0,(j-1/2)^2+(k-1/2)^2) \nonumber \\ & &
- \left[ (j-1/2)^2+(k-1/2)^2 \right] 
z_{5/2}(0,(j-1/2)^2+(k-1/2)^2) \Big) \nonumber \\ & &
+ \sum_{j=1}^{\infty} \left[ j^4 z_{7/2}((\pi\eta)^2/2,j^2) 
- j^2 z_{5/2}((\pi\eta)^2/2,j^2) \right]
+ \sum_{j=1}^{\infty} (-1)^j \left[ j^4 z_{7/2}((\pi\eta)^2/2,j^2) 
- j^2 z_{5/2}((\pi\eta)^2/2,j^2) \right] \nonumber \\ & &
+ 2 \sum_{j,k=1}^{\infty} (-1)^j  \left[ (j^2-k^2)^2 z_{7/2}((\pi\eta)^2/2,j^2+k^2) 
- (j^2+k^2) z_{5/2}((\pi\eta)^2/2,j^2+k^2) \right] \nonumber \\ & &
+ \sum_{j=1}^{\infty} \left[ (j-1/2)^4 z_{7/2}(0,\eta^2/2+(j-1/2)^2) 
- (j-1/2)^2 z_{5/2}(0,\eta^2/2+(j-1/2)^2) \right] \nonumber \\ & &
+ 2 \sum_{j,k=1}^{\infty} \Big( \left[ (j-1/2)^2-k^2 \right]^2
z_{7/2}(0,\eta^2/2+(j-1/2)^2+k^2) \nonumber \\ & &
- \left[ (j-1/2)^2 +k^2 \right] z_{5/2}(0,\eta^2/2+(j-1/2)^2+k^2) \Big) \Bigg\} .
\end{eqnarray}


\begin{thebibliography}{10}
\bibitem{rque50}
This value is fully consistent with that reported in Refs.\cite{Goldoni96,Bonsall} 
In addition, it also provides us with
the Madelung constant of the hexagonal Wigner crystal on a single 
charged plate $M=-E(\sqrt{3},0)/(e^2 \sqrt{n \pi}) \simeq 1.106103$,
as can be found in Refs. \cite{Levin02,Shklovskii}.
\end{thebibliography}
\end{document}